\documentclass[twocolumn, prb,showpacs,superscriptaddress]{revtex4-1}

\usepackage{amsmath,amssymb}
\usepackage{amsmath}
\usepackage{braket}
\usepackage{graphicx}
\usepackage{subfigure}
\usepackage{color}
\newcommand{\Ham}{\ensuremath{\mathcal{H}}}
\newcommand{\dg}{\dagger}
\newcommand{\pd}{{\phantom{\dagger}}}

\usepackage[draft]{fixme} 

\newcommand{\Omegaat}{\Omega_{\text{at}}}

\begin{document}
\title{Single polaron properties for double-well electron-phonon coupling}

\author{Clemens P.J. \surname{Adolphs}}
\email{cadolphs@phas.ubc.ca}
\affiliation{\!Department \!of \!Physics and Astronomy, \!University of\!
  British Columbia, \!Vancouver, British \!Columbia,\! Canada,\! V6T \!1Z1}

\author{Mona Berciu}
\affiliation{\!Department \!of \!Physics and Astronomy, \!University of\!
  British Columbia, \!Vancouver, British \!Columbia,\! Canada,\! V6T \!1Z1}
\affiliation{\!Quantum Matter \!Institute, \!University of British Columbia,
  \!Vancouver, British \!Columbia, \!Canada, \!V6T \!1Z4}

\date{\today}

\begin{abstract}
We show that in crystals where light ions are symmetrically intercalated
between heavy ions, the electron-phonon coupling for carriers located
at the light sites cannot be described by a Holstein model. We
introduce the double-well electron-phonon coupling model to describe
the most interesting parameter regime in such systems, and study it in the
single carrier limit using the momentum average approximation. For
sufficiently strong coupling, a  small polaron with a
robust phonon cloud appears at low energies. While some of its
properties are
similar to those of a Holstein polaron, we highlight some crucial
differences. These prove that the physics of the
double-well electron-phonon coupling model cannot be reproduced with a
linear Holstein model.
\end{abstract}


\maketitle

\section{Introduction}
When charge carriers couple to phonons, magnons, or other bosonic
excitations, the resulting dressed quasiparticles -- the
polarons -- often behave drastically different from the free carriers.
This is why understanding the consequences of carrier-boson coupling
is important for many materials such as organic
semiconductors,\cite{organic_1, organic_2} cuprates,\cite{cuprates_1,
  cuprates_2, cuprates_3, cuprates_4, cuprates_5, cuprates_6}
manganites,\cite{manganites} two-gap superconductors like
MgB$_2$,\cite{mgb_1, mgb_2, mgb_3, mgb_4} and many more. To describe
them, many models of varying complexity have been devised
and studied. The simplest is the Holstein model for electron-phonon
coupling,\cite{holstein}
 where carriers couple to a
branch of dispersionless optical phonons through a momentum-independent
coupling  $g$. Physically, it describes a modulation of the
on-site potential of the carrier due to the deformation of the
``molecule'' hosting it. Longer-range coupling that modulates
the carrier's on-site potential leads to  $g(q)$ couplings
that depend on the boson's momentum, such as the
Fr\"ohlich\cite{frohlich} or the breathing-mode models.\cite{bm} If the
bosons modulate the hopping of the carrier,  the
coupling $g(k,q)$ depends on the momenta of both carrier and
boson, as is the case in the Su-Schrieffer-Heger (SSH)
model\cite{SSH,extra} or for a hole coupled to magnons  in an
antiferromagnet, as described by a $tJ$ model.\cite{tJmod}

All these electron-phonon coupling models assume that the coupling is
\emph{linear} in the lattice displacements. This is a natural
assumption because if the displacements are small, the linear term is
the most important contribution. However, the coefficient of the
linear term may vanish due to symmetries of the crystal. In such
cases, the most important contribution is the {\em quadratic} term.

Here we introduce, motivate and study in detail a Hamiltonian
describing such quadratic electron-phonon (e-ph) coupling relevant for
many common crystal structures, consisting of intercalated sublattices
of heavy and light atoms.  We focus on the single carrier limit and
the parameter regime where the carrier dynamically changes the
effective lattice potential from a single-well to a double-well;
hence, we call this {\em the double-well e-ph coupling}.  We use the
momentum-average approximation\cite{MA_berciu,MA_goodvin} to compute
the properties of the resulting polaron with high accuracy. We find
that although the polaron shares some similarities with the Holstein
polaron, it also differs in important aspects. Indeed, we show
that the physics of the double-well e-ph coupling model cannot be
described by a renormalized linear Holstein model.

To the best of our knowledge, this is the first systematic,
non-perturbative study of such a quadratic
model. Previously\cite{adolphs_nonlinear} we studied the effect of
quadratic (and higher) corrections added to a linear term.  Weak,
purely quadratic coupling was studied using perturbation theory in
Refs. \onlinecite{quadratic_1}, \onlinecite{quadratic_2}. Other works
considered complicated non-linear lattice potentials and couplings but
treated the oscillators classically,\cite{anharmonic_Holstein,
  kenkre1, kenkre2} or discussed anharmonic lattice potentials but for
purely linear coupling.\cite{anharmonic1, anharmonic2} Away from the
single-carrier limit, the Holstein-Hubbard model in infinite
dimensions was shown to have parameter regions where the effective
lattice potential has a double-well shape;\cite{infinite_1,
  infinite_2, infinite_3} this was then used to explain
ferroelectricity in some rare-earth oxides.\cite{[{See }] [{ and
      related references}] double_well_2} However, the effect of a
double-well e-ph coupling on the properties of a single polaron were
not explored in a fully quantum-mechanical model on a low-dimensional
lattice.

This work is organized as follows: in Section~\ref{sec:model} we
introduce the Hamiltonian,  motivate its use for relevant systems, and
discuss all approximations made in deriving it. In
Section~\ref{sec:formalism} we review the theoretical means by which
we study our Hamiltonian. In Section~\ref{sec:results} we present our
results, and in Section~\ref{sec:summary} we give our concluding
discussion and an outlook for future work.

\section{The model}\label{sec:model}
\begin{figure}[t]
\subfigure[\label{fig:1d_chain}]{\includegraphics[width=0.35\textwidth]{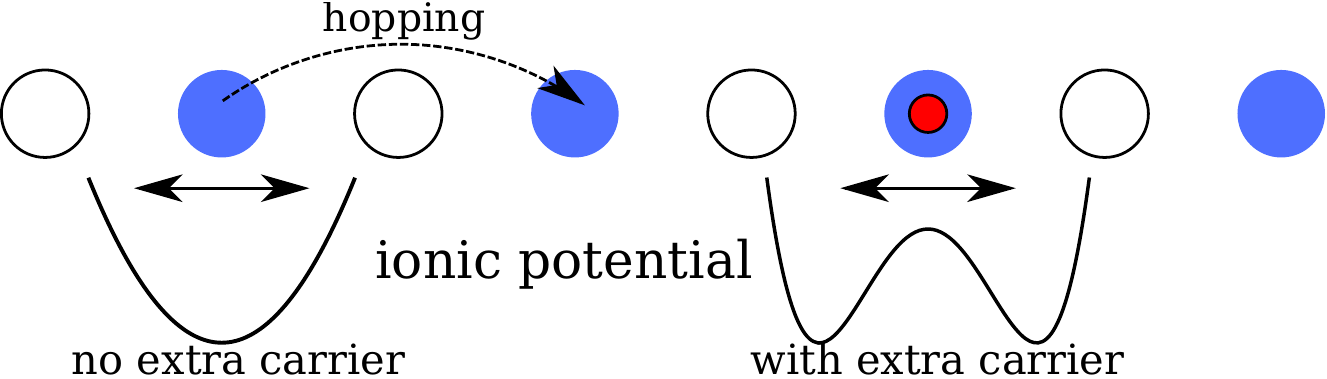}}
\vspace{5mm}

\subfigure[\label{fig:2d_lattice}]{\includegraphics[width=0.2\textwidth]{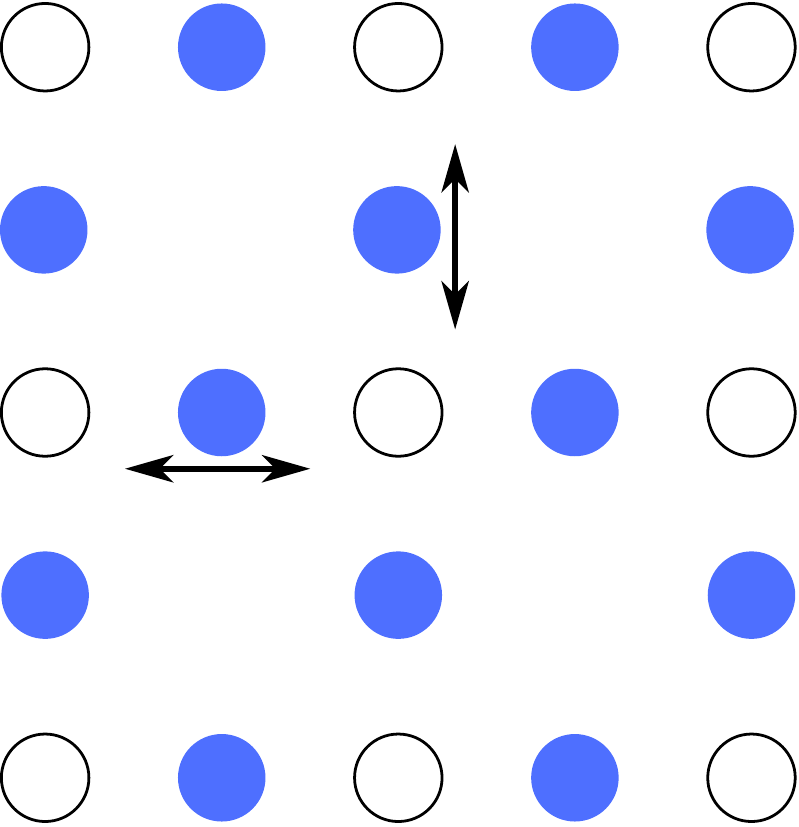}}
\caption{(color online) Sketch of the crystal structures discussed in
  this work: (a) 1D chain, and (b) 2D plane, consisting of light
  atoms (filled circles) intercalated between heavy atoms (empty
  circles).  In the absence of carriers, the ionic potential of a
  light atom is a simple harmonic well. In the presence of a carrier,
  the ionic potential of the light atom hosting it remains an even
  function of its longitudinal displacement, so the linear e-ph
  coupling vanishes. In suitable conditions the effective ionic
  potential becomes a double well (see text for more details).  }
\end{figure}

The crystal structures of interest are illustrated in
Fig.~\ref{fig:1d_chain} for 1D, and Fig.~\ref{fig:2d_lattice} for 2D
cases. The 3D crystal would have a perovskite structure but we do not
discuss it explicitly because, as we show below, dimensionality plays no
role in determining the polaron properties.

The undoped compound is an insulator made of
light atoms, shown as filled circles, intercalated between heavy ones,
shown as empty circles. To zeroth order, the vibrations of the heavy
atoms can be ignored while those of the light atoms are described by
independent harmonic oscillators $\Ham_{\text{ph}}=\Omega \sum_i
b_i^\dg b_i^\pd$, where $b_i$ annihilates a phonon at the $i^{th}$
light atom. (We set the mass of the light ions $M=1$, and also
$\hbar=1$). In reality there is weak coupling between these
oscillators giving rise to a dispersive optical phonon
branch. However, the dispersion can be ignored if its bandwidth is small compared
to all other energy scales. We do so in the following.

Consider now the addition of a carrier. If it occupies orbitals
centered on the heavy atoms, its coupling to the oscillations of the
light atoms is described by breathing-mode coupling models.\cite{bm}
Here we are instead interested in the case where the carrier is
located on the light atoms. Such is the situation for a CuO$_2$ plane
as shown in Fig.~\ref{fig:2d_lattice}, since the parent compound is a
charge-transfer insulator\cite{ZSA} so that upon doping, the holes
reside on the light O sites (of course, there are
additional complications due to the magnetic order of the Cu spins; we
ignore these degrees of freedom in the following).  The carrier moves
through nearest-neighbor hopping between light atoms: $\hat
T=-t\sum_{\langle i,j \rangle} \left(c_i^\dg c_j^\pd + h.c.\right)$,
where $c_i$ is the carrier annihilation operator at light atom $i$.

Given the symmetric equilibrium location of the light ion hosting the
carrier between two heavy ions, it is clear that the e-ph coupling
cannot be linear in the displacement $\delta x_i$ of that light ion:
the sign of the displacement cannot matter. Thus, e-ph coupling in
such a material is not described by a Holstein model. This assertion
is supported by detailed modelling. For simplicity, we assume that the
interactions with the neighboring heavy atoms are dominant
(longer-range interactions can be easily included but lead to no
qualitative changes). There are, then, two distinct contributions to
the e-ph coupling:

\paragraph{Electrostatic coupling:} The carrier changes the
total charge of the light ion it resides on. If the distance between
adjacent light and heavy ions is $d$, and if $U(x)$ is their
additional Coulomb interaction due to the carrier, then the potential
increases by $U(d+\delta x_i) + U(d-\delta x_i)$. This is an even
function and thus has no linear (or any odd) terms in $\delta x_i$.
The coefficient of the quadratic term $(\delta x_i)^2$ can be either
positive or negative, depending on the charge of the carrier (electron
or hole).

\paragraph{Hybridization:} Even though charge transport
is assumed to take place in a light atom band, there is always some
hybridization $t_{lh}$ allowing the carrier to hop onto an adjacent
heavy ion. If $\Delta$ is the corresponding energy increase, assumed
to be large, then the carrier can lower its on-site energy by
$-t_{lh}^2/\Delta$ through virtual hopping to a nearby heavy ion and
back. The hopping $t_{lh}$ depends on the distance between ions; for
small displacements $t_{lh}(\delta x) \approx t_{lh} (1 + \alpha
\delta x)$ where $\alpha $ is some material-specific constant. Because
the light ion is centered between two heavy ions, such contributions
add to
$\frac{-t_{lh}^2}{\Delta}\left[ (1+\alpha \delta x)^2 + (1-\alpha
  \delta x)^2\right] = \frac{-2t_{lh}^2}{\Delta}\left[ 1 + \alpha^2
  (\delta x)^2 \right].
$
The potential is again even in $\delta x$.  In this case, the coefficient of the
quadratic term is always negative.

Given that $\delta x_i \sim b_i + b_i^\dg$, it follows that the
largest (quadratic)  contribution to the e-ph coupling for such a
crystal has the general form:
\begin{equation*}
  \Ham^{(2)}_{\text{e-ph}} = g_2\sum_{i}^{} c_i^\dg c_i \left( b_i + b_i^\dg\right)^2
\end{equation*}
where all prefactors have been absorbed into the energy scale
$g_2$, and the sum is over all light ions. From the analysis above
we know that $g_2$ may have
either sign.

Physically, $\Ham^{(2)}_{\text{e-ph}}$ shows that the presence of a
carrier modifies the curvature of its ion's lattice potential, and
thus changes the phonon frequency at that site from $\Omega$ to
$\Omega_{\text{at}} = \sqrt{\Omega^2 + 4\Omega g_2}$. If $g_2 > 0$
then $\Omegaat > \Omega$, making phonon creation more costly. As we
show in Appendix \ref{newap}, this leads to a rather uninteresting
large polaron with very weakly renormalized properties. This is why in
the following we focus on the case with $g_2 < 0$.

For sufficiently negative $g_2$, $\Omegaat$ vanishes or becomes
imaginary, {\em i.e.} the lattice is unstable. This is unphysical; in
reality the bare ionic potential contains higher order terms that
stabilize the lattice. This means that for $g_2 < 0$ we must include
anharmonic (quartic) terms in the phonon Hamiltonian and, for
consistency, also in the e-ph coupling, so that
\begin{eqnarray*}
\Ham_{\text{ph}} &&= \Omega \sum_i b_i^\dg b_i^\pd + \Theta \sum_i
(b_i^\dg + b_i^\pd)^4\\
\Ham^{(4)}_{\text{el-ph}}  &&=
\sum_{n\in\{2,4\}} g_n \sum_i c_i^\dg c_i^\pd ( b_i^\dg + b_i^\pd
)^n,
\end{eqnarray*}
where $\Theta$ is the scale of the anharmonic corrections. In
physical situations $\Theta \ll \Omega$ and $0< g_4 \ll |g_2|$, or
the Taylor expansions would not be sensible starting points.

\begin{figure}[t]
  \centering
  \includegraphics[width=0.48\textwidth]{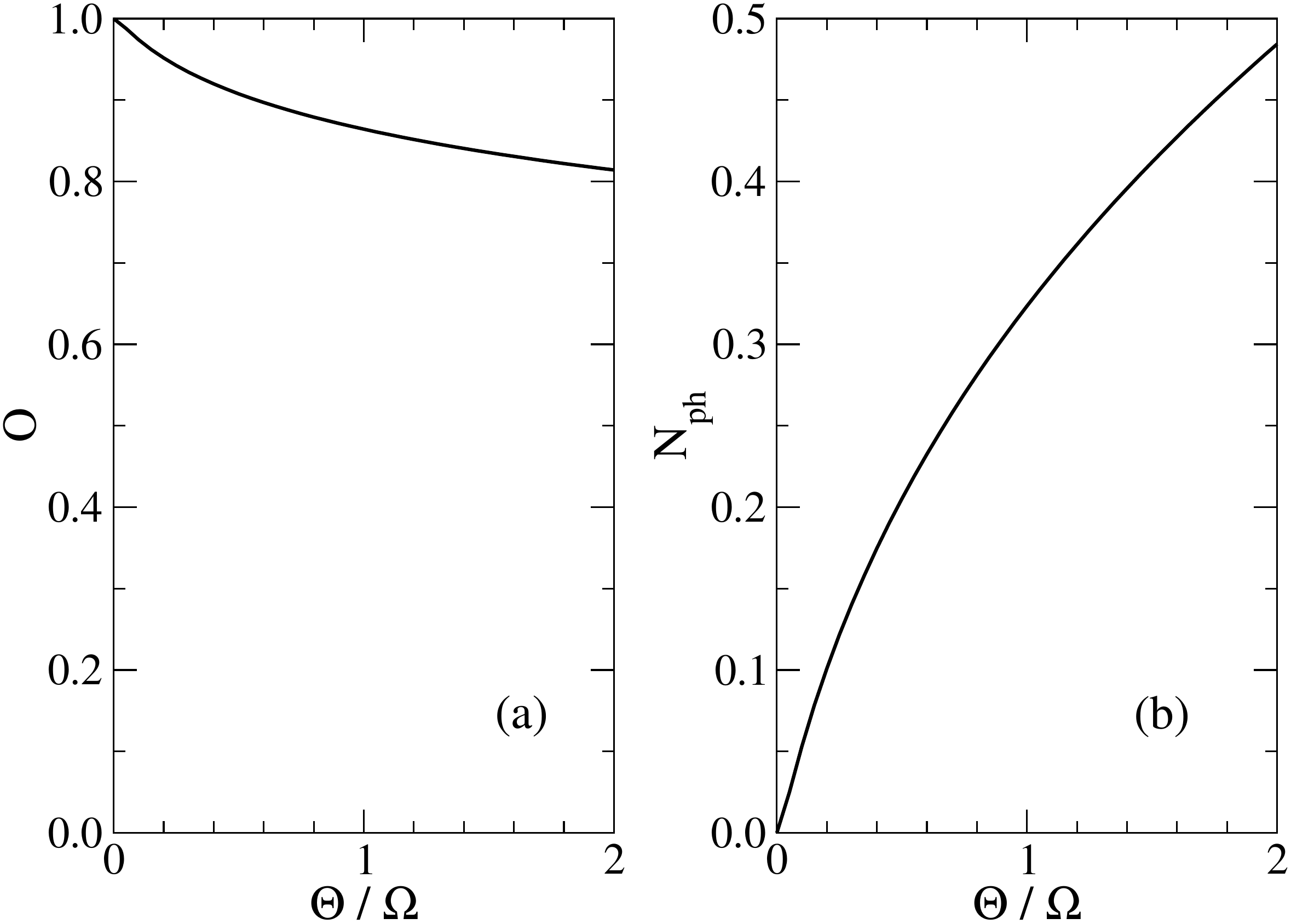}
  \caption{(a) Overlap
    between the
    undoped ground-states with and without anharmonic corrections, and (b)
    the average number of phonons per site in the undoped system, due
    to anharmonic
    corrections, as a function of $\theta/\Omega$. }
  \label{fig:quartic_effects}
\end{figure}

The anharmonic terms in $\Ham_{\text{ph}}$ make the
total Hamiltonian unwieldy, because the phonon vacuum $|0\rangle$ is
no longer the undoped ground-state, and the new undoped ground state
$\ket{\tilde 0}$ has no simple analytical expression.
In order to be able to proceed with an analytical approximation, we
argue that these terms can be absorbed into the e-ph
coupling; this is a key approximation of the model. The reasoning is
as follows:
At those lattice sites that do not have a carrier, the quartic terms
have little effect if $\theta \ll \Omega$.  This statement is verified
by exact diagonalization of $\Ham_{\text{ph}}$. Results are shown
in Fig.~\ref{fig:quartic_effects} where we plot the overlap
$O=|\langle 0 |{\tilde 0}\rangle|^2$ (per site) between the undoped
ground-states
with and without anharmonic corrections, as well as the average number
of phonons at a site of the undoped lattice. Even for unphysically
large values  $\Theta/\Omega\sim 1$, the overlap $O$ remains close to 1
while $N_{\text{ph}} \ll1$, showing that the undoped ground-state has
not changed significantly in the presence of anharmonic
corrections. From now we ignore these corrections at sites without an
additional carrier.

However, for sites that have a carrier present,
we cannot ignore the anharmonic term: As discussed, it
is crucial for stabilizing the lattice. Since this term is similar to the
quartic term in the e-ph coupling, they can both be grouped
together, resulting in the approximate
Hamiltonian for our crystal:
\begin{multline}
  \label{eq:H}
  \Ham = \hat T + \Omega \sum_i b_i^\dg b_i^\pd +
  g_2 \sum_i c_i^\dg c_i^\pd \left( b_i^\dg + b_i^\pd\right)^2
\\ + (g_4 + \Theta) \sum_i c_i^\dg c_i^\pd \left( b_i^\dg + b_i^\pd \right)^4
\end{multline}
with an effective quartic e-ph coupling term $g_4 +
\Theta$, which from
now on we will simply call $g_4$. This is the Hamiltonian that
we investigate in this work.

Before proceeding, let us review what we are neglecting when we
discard the anharmonic corrections at the unoccupied sites. Besides
ignoring the change in the undoped ground state from $|0\rangle $ to
$|{\tilde 0}\rangle$ (which is a reasonable approximation if
$\theta/\Omega \ll 1$, as discussed), we also assume that only the
e-ph coupling can change the number of phonons in the system, whereas in
the full model the phonon number is also changed by anharmonic
corrections. This latter approximation is valid if the timescale for
anharmonic phonon processes $\tau_{4} \sim 1/\Theta$ is much longer than
the characteristic polaron timescale $\tau_p \sim t m / m^*$, where
$m^*$ is the effective polaron mass.

Let us briefly summarize the basic properties of the lattice
potential, which equals  $V_e(\delta x) = \Omega^2 (\delta x)^2/2$ for sites
without an extra
carrier, and  $V_c(\delta x) =
\Omegaat^2 (\delta x)^2/2 + 4 \Omega^2
g_4 (\delta x)^4$
for sites with one carrier.
If $g_2 > -\Omega/4$, the first term describes a harmonic
well with frequency $\Omegaat$ and  $V_c(\delta x) $
describes a single well centered at $\delta x = 0$. If $g_2< -\Omega/4$,
however,  $\Omegaat$ becomes purely imaginary. In this case,
$V_c(\delta x) $  becomes a double-well potential with a
local maximum at  $\delta x = 0$.  The two wells are centered at
$
  \pm x_{\text{eq}} = \pm \sqrt{ \frac{-\Omega - 4g_2}{16 \Omega g_4}}.
$ For $\delta x \approx \pm x_{\text{eq}}$ we obtain
 $ V_c(\delta x) \approx V(x_{\text{eq}}) - \Omegaat^2 (\delta x \mp
  x_{\text{eq}})^2$,
which locally describes a harmonic well of frequency $\Omega_{\text{eff}}^2 =
-2\Omegaat^2$. Interestingly, this is independent of $g_4$, whose only
role is to control the location and depth of the two wells
(they are further apart and deeper for \emph{smaller}
$g_4$).

\section{Formalism}\label{sec:formalism}

We want to find the single particle Green's function
$G(k, \omega) = \braket{ 0 | c_k^\pd \hat G(\omega) c_k^\dg | 0}$,
where $\hat G(\omega) = [\omega - \Ham + i\eta]^{-1}$ is the resolvent
of Hamiltonian \eqref{eq:H}. From this, we can obtain all the
polaron's ground state properties as well as its
dispersion.\citep{MA_goodvin}

Grouping
terms in the Hamiltonian according to how they affect the phonon number, we
rewrite
$\Ham = \Ham_0 + \Ham_p + \Ham_2 + \Ham_4$ with $\Ham_0 = \hat T +
\Omega \sum_i b_i^\dg b_i + g_2 + 3g_4$ and $\Ham_p =
  \sum_i n_i b_i^\dg b_i (2g_2 + 6g_4 + 6g_4 b_i^\dg b_i)$ do not change
  the number of phonons, while $
 \Ham_2 = \sum_i n_i \left[(g_2 + 6g_4)(b_i^{\dg, 2}\! + b_i^2) +
        4g_4 (b_i^{\dg,3} b_i + b_i^\dg b_i^3)\right] $ and
$ \Ham_4 = g_4 \sum_i n_i \left( b_i^{\dg, 4} + b_i^4\right)$ change
 it by $\pm2$ and $\pm4$, respectively.
The constant  $g_2 + 3 g_4$ in $ \Ham_0$ is absorbed into $\omega$ in the following derivations, but plots of the spectral weight will show actual energies.

One important property of this Hamiltonian is that it preserves the
phonon number parity on each site: because its terms only change the
number of phonons by multiples of two, any eigenstate is a sum of
basis states having only even (or only odd) number of phonons.  The
Hilbert space can thus be divided into an \textit{even} and an
\textit{odd} (phonon number) sector, which can be diagonalized
separately. We  emphasize that this symmetry is
  different from the parity symmetry under a
  global lattice inversion $\vec{r} \rightarrow -\vec{r}$. The latter has
  been studied extensively for the linear Holstein
  model,\cite{linear_symmetry} where it was shown that polaron states
  with total momentum $K = 0, \pi$ have well defined (spatial) parity.
  The phonon number parity, on the other hand, corresponds to a
  unitary transformation $b_i^\dg \rightarrow -b_i^\dg$, i.e., a local
  inversion of the phonon coordinates.  The number parity symmetry
  also correlates with the \emph{local} spatial parity of the ions,
  since the spatial parity operator for site $i$ can be written as $\hat P_i =
  \exp(i\pi b_i^\dg b_i)$.

\subsection{The even sector}
We compute the Green's function via the same continued matrix fractions
method\cite{mirko_efficient} previously  used by us to compute the
Green's function of a generalized Holstein model with linear and higher-order
terms\cite{adolphs_nonlinear} within the framework of the momentum
average (MA) approximation. This approximation was shown to be highly
accurate for models with Holstein
coupling.\cite{MA_berciu,MA_goodvin} The reasons for this (such as
obeying exact sum rules) can be verified to hold for this model, too.
  To be specific, here we implement the  MA$^{(2)}$ flavor
  which allows us to also locate the continuum lying above the
  polaron band.\cite{MA1}

We begin our derivation by dividing the
Hamiltonian into
$\Ham = \Ham_0 + \Ham_1$ with $\Ham_1 = \Ham_p + \Ham_2 + \Ham_4$.
Using Dyson's identity $\hat G(\omega) =\hat G_0(\omega) +\hat
G(\omega)\Ham_1 \hat G_0(\omega)$, where  $\hat G_0(\omega) = [\omega
  - \Ham_0 + i\eta]^{-1}$,
we obtain
\begin{multline}
\label{geq}
  G(k,\omega) = G_0(k,\omega) \Big[
    1 + \\ \sum_i \frac{e^{ikR_i}}{\sqrt{N}}
    (g_2 + 6g_4) F_1(k,\omega;i,i)
    +  g_4 F_2(k,\omega;i,i)\Big]
\end{multline}
with
$F_n(k,\omega;i,j) = \braket{0 | c_k \hat G(\omega) c_i
    (b_i^\dg)^{2n-2} (b_j^\dg)^2 |0}$
being the generalized propagator for a system with $2n$ phonons in total, $2n-2$
of them on site $i$ with the other two on site $j$. The difference
between MA$^{(2)}$
and the original MA, which we also call MA$^{(0)}$, is
that for $F_1$ we also use its exact equation of motion (EOM),
\begin{widetext}
  \begin{multline}
    F_1(k,\omega;i,j) = G(k,\omega;j)G_0(j-i,\omega-2\Omega)(2g_2 +
    12g_4)
    + F_1(k,\omega;j,j)G_0(j-i,\omega-2\Omega)(4g_2 + 36g_4)  \\
  + 8g_4  F_2(k,\omega;j,j)G_0(j-i,\omega-2\Omega)+ \sum_l
  G_0(l-i,\omega-2\Omega)\left[
      F_2(k,\omega;l,j)(g_2 + 6g_4) + F_3(k,\omega;l,j) g_4\right].
	\label{eq:eom_F1}
  \end{multline}
which is obtained by applying Dyson's identity again, and introducing
$G(k,\omega; j) = \braket{0 | c_k^\pd \hat G(\omega) c_j^\dg | 0}$
and $G_0(j-i,\omega) = \braket{0 | c_j^\pd \hat G_0(\omega) c_i^\dg | 0}$.
The equations of motion for the $F_n$ propagators with $n \ge 2$ are
approximated
by replacing the free propagator $G_0(j-i, \omega-2n \Omega)
\rightarrow \delta_{i,j} \bar g_0(\omega - 2n\Omega)$, where $\bar
g_0(\omega)={1\over N} \sum_{k}^{} G_0(k,\omega)$ is the momentum
averaged free propagator.  At low
energies this is a good approximation because $G_0(j-i, \omega-2n
\Omega)$ decays exponentially with the distance $|j-i|$ if $\omega -
2n\Omega < - 2dt$ in $d$ dimensions. This is also justified by the
variational meaning of the MA approximations, discussed at
length elsewhere.\cite{bar,MA1} (Basically, MA$^{(2)}$  assumes that all
phonons in the cloud are at the same site but also allows for a pair
of phonons to be created at a site away from the cloud).

The resulting
EOMs are different
depending on whether $i = j$ or
$i \not= j$.
If we define $F_n^=(k,\omega;i) = F_n(k,\omega;i,i)$ and
$F_n^{\neq}(k,\omega;i,j) = F_n(k,\omega;i,j)$ for $i\ne j$, we obtain
  \begin{multline}
    F_n^=(k,\omega;i) = \bar g_0(\omega - 2n\Omega)\Big[ F_{n-2}^= (2n)^{\bar 4} g_4 +
      F_{n-1}^= \left( (g_2 + 6g_4)(2n)^{\bar 2} + 4g_4 (2n)^{\bar 3}\right) + \\
      (4n g_2 + 12 n g_4 + 24n^2 g_4) F_n^= +
      (g_2 + 6g_4 + 8ng_4) F_{n+1}^= + g_4 F_{n+2}^= \Big].
      	\label{eq:Fn_eq}
    \end{multline}
    \begin{multline}
    F_n^{\neq}(k,\omega;i,j) = \bar g_0(\omega - 2n\Omega)\Big[g_4 (2n-2)^{\bar 4} F_{n-2}^{\neq} + \left((g_2 + 6g_4)(2n-2)^{\bar 2}
    + (2n-2)^{\bar 3} \cdot 4g_4\right) F_{n-1}^{\neq} +\\
  \left[ 2(2n-2)g_2 + 12(n-1)g_4 + 6(2n-2)^2 g_4 \right] F_n^{\neq} + \left[ g_2 + 6g_4 + 4(2n-2) g_4\right] F_{n+1}^{\neq}
  + g_4 F_{n+2}^{\neq} \Big]
  		\label{eq:Fn_neq}
    \end{multline}
\end{widetext}
where we use the notation $x^{\bar n}=x! / (x-n)!$. We also
omitted the arguments from the $F_n$ appearing on the right hand
sides,  as they remain unchanged.

These EOMs connect generalized Green's functions $F_n$ with $F_{n\pm 1}$ and
$F_{n\pm 2}$.
We reduce this to a first order
recurrence relation\cite{adolphs_nonlinear} by introducing vectors
$W_n^= = (F_{2n}^=, F_{2n+1}^=)$ and
analogously for $W_n^{\neq}$.  Below, we write
$W_n$ without the index $=$ or
$\neq$ for results that apply to both $W_n^=$ and $W_n^{\neq}$.  By inserting
the EOMs into the definition of $W_n$, we obtain a matrix EOM for the $W_n$,
\begin{equation}
  \gamma_n W_n = \alpha_n W_{n-1} + \beta_n W_{n+1}.
\label{w}
\end{equation}
The coefficients of these matrices are read off from the EOM for the
$F_n$. They are listed in appendix \ref{sec:appendix_even_matrices}.

Using the fact that $\lim_{n\rightarrow \infty} A_n = 0$ we can
show\cite{adolphs_nonlinear} that $W_n = A_n
W_{n-1}$ with
 $ A_n = \left[ \gamma_n - \beta_n A_{n+1} \right]^{-1} \alpha_n$.
By introducing a suitably large cut-off $N$ where we set $A_{N+1} = 0$, we
can compute all $A_n$ with $n\le N$ as
continued matrix fractions. Knowledge of $A_1$ allows us to express
$F_2$ and $F_3$ in terms of
$F_1$ and $F_0 = G$. Following a series of steps presented in appendix
\ref{sec:appendix_even_EOMs}, we obtain a closed equation for $F_1$ in
terms of $G$, which we then finally use to compute $G$. The end result
of these manipulations is the self energy
\begin{equation*}
  \Sigma(\omega) = \frac{(g_2 + 6g_4 + A_1^=|_{12} g_4) \tilde g_0(\omega)
    a_0^=}{1-\tilde g_0(\omega)(a_1^= - a^{\neq})} + g_4A_1^=|_{11}.
\end{equation*}
with $\tilde g_0(\omega) = \bar g_0(\omega - 2\Omega - a^{\neq}$) and
the other coefficients defined in appendix
\ref{sec:appendix_even_EOMs}. The independence of the self-energy on
momentum is the consequence of the local form of the coupling and of the
non-dispersive phonons, similar to the MA results for the Holstein
model.\cite{MA1} Momentum-dependence would be acquired in a higher
flavor of MA, but is likely to be weak. Finally, the Green's function is:
\begin{equation}
G(k, \omega) = \frac{1}{\omega - \epsilon_k - \Sigma(\omega) + i\eta}.
\end{equation}
One can now use the matrices $A_n$ to
  generate the generalized propagators $F_n$, which allow one to
  reconstruct the entire polaron wavefunction (within this
  variational space).\cite{Can} For the
  quantities of interest here, however, the single-particle Green's
  function suffices.

\subsection{The odd sector}
Here we calculate the
Green's function for a state that already has a phonon in the
system. Since the phonon number can only change by $2$ or $4$, this single
phonon can never be moved to another site, so it is natural to compute
the Green's function in real space.
The most general such real space Green's function is:
\begin{equation*}
  G_{ijl}(\omega) = \braket{0| b_l^\pd c_j^\pd \hat G(\omega) c_i^\dg b_l^\dg | 0}.
\end{equation*}
Applying the Dyson identity leads to the EOM
\begin{multline*}
  G_{ijl}(\omega) = G_{0}(j-i,\omega - \Omega) \\
+ \sum_{i'} G_{0}(i'-i,\omega-\Omega) \braket{0 | b_l^\pd
  c_j^\pd \hat G(\omega) \Ham_1 c_{i'}^\dg b_l^\dg | 0}.
\end{multline*}
We then split the sum over all lattice sites into a term $i'=l$ where
the electron is on the same site as the extra phonon, and a sum over
all the other sites. The subsequent steps are very similar to those
for the even-sector Green's function. We summarize them in Appendix
\ref{sec:appendix_odd}, where we also discuss how various
  propagators that enforce translational symmetry -- \textit{i.e.}
  propagators defined in momentum space -- can be obtained from these
  real-space Green's functions.

The end result for the real-space Green's functions is
$G_{ijl}(\omega) =  G_0(j-i,\tilde \omega)
+ G_0(l-i,\tilde
    \omega)G_0(j-l,\tilde \omega)(a^=_{\text{o}} -
    a^{\neq}_{\text{o}})[1 - \bar g_0(\tilde \omega)(a^=_{\text{o}} -
  a^{\neq}_{\text{o}})]^{-1}$
where $\tilde \omega = \omega - a^{\neq}_{\text{o}} - \Omega$.
The coefficients $a^=_o$ and $a^{\neq}_o$ are listed in appendix
\ref{sec:appendix_odd}.

\section{Results}\label{sec:results}

\subsection{Atomic limit: $t=0$}

We begin our analysis with the atomic limit since it is a good
starting point for understanding the properties of the small
polaron, which is the more interesting regime. However, we note an
important distinction between the Holstein model and our double-well
model. In the former, the atomic limit is the infinite-coupling
limit. In the latter, $g_4$ sets an additional energy scale. Thus, the
atomic limit is not the same as the strong coupling limit; the latter
also requires that $g_4/|g_2|$ be small.

Before doing any computations, we can describe some general features of
the spectrum.  As already discussed, the phonon
component of the wavefunctions has either even or odd phonon number
parity. Since this is due  to the spatial
symmetry in the local ionic displacement, in any
eigenstate the ion is equally likely to be found in either well. As
usual, the ground state has \emph{even} symmetry since it has no nodes
in its wavefunction. Subsequent eigenstates always have one more node
than the preceding eigenstate, so states with even and odd parity
alternate. The exception is the
limit of infinite well separation, $g_4/|g_2| \rightarrow 0^+$, where the
$2n^{th}$ and $2n+1^{st}$ eigenstates become degenerate. The system
can then spontaneously break parity to have the ion definitely located
in the left or in the right well, like in a ferroelectric. For a
finite $g_4$ this is not possible in the single carrier
limit, but it can be achieved at finite carrier concentration through spontaneous
symmetry breaking.

As discussed, our results are obtained with MA. In the
atomic limit MA is exact\cite{MA_berciu} because for $t=0$ the free
propagator is diagonal in real-space so the terms ignored by MA
vanish. Thus, MA
results must be identical here to those obtained by other exact
means. To check our implementation of MA, we used exact
diagonalization (ED) with up
to a few thousand phonons; this suffices for an accurate
computation of the  first few eigenstates.
ED and MA results agree, as required.

\begin{figure}[t]
  \centering
  \subfigure{\includegraphics[width=0.48\textwidth]{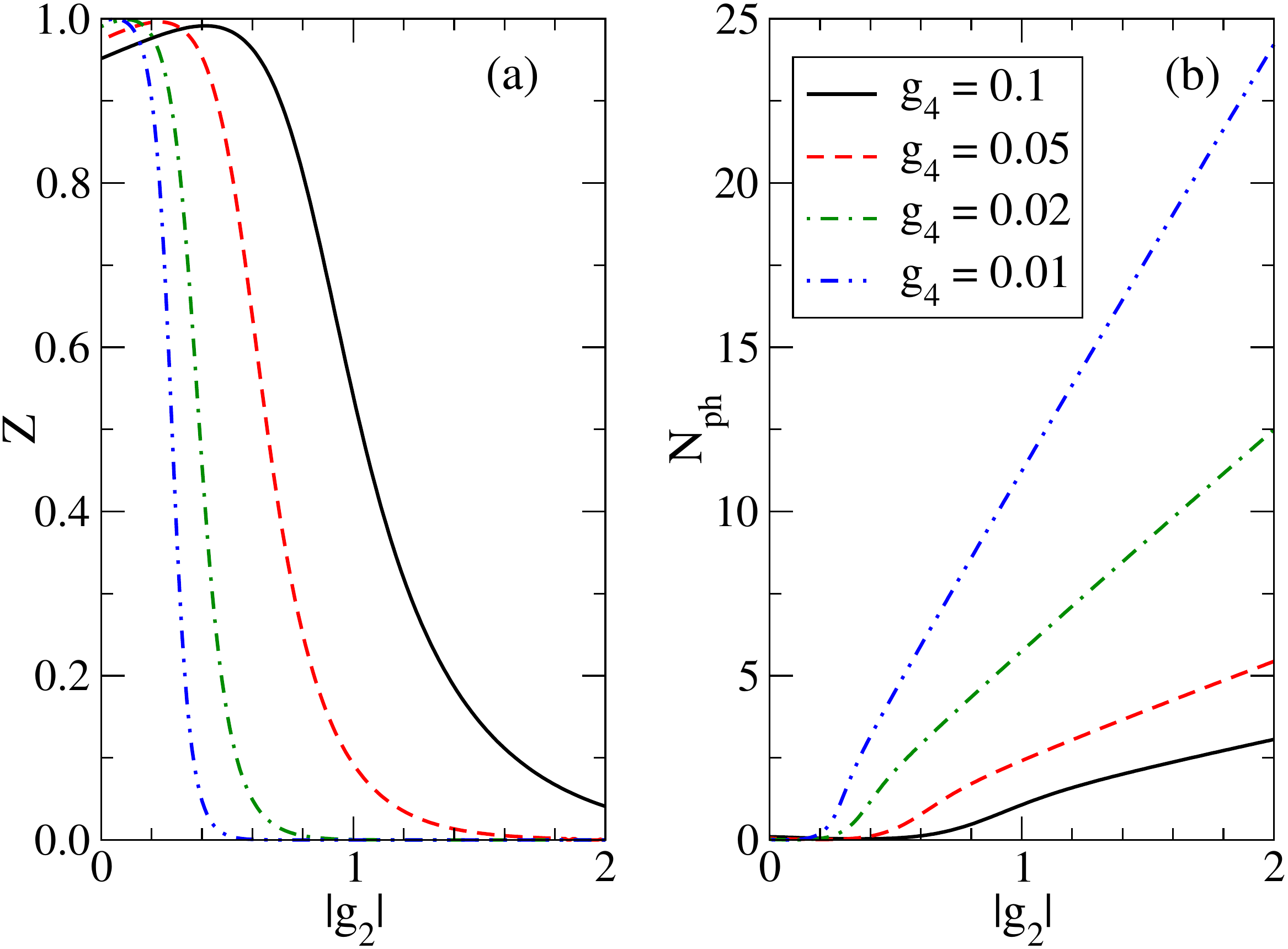}}
  \caption{(color online) Polaron ground-state properties  in the
    atomic limit, for several values of the
    $g_4$: a) quasiparticle weight, and b) average number of
    phonons in the phonon cloud. Other parameters are $\Omega = 0.5, t=0$.}
  \label{fig:atomic_limit_polaron}
\end{figure}

Figure~\ref{fig:atomic_limit_polaron} shows the ground-state
quasiparticle weight $Z$ (the overlap between the polaron
ground-state and the non-interacting carrier ground-state),  and the
ground-state average number of phonons in the cloud,
$N_{\text{ph}}$, as a function of $g_2<0$, for various values of
$g_4$. $Z$ has an interesting behavior. At $g_2 = 0$ it is slightly
below $1$ because of the quartic terms. As $|g_2|$ is increased, $Z$
first rises towards a value close to $1$ and then sharply drops. This
turnaround is caused by the terms that involve both $g_2$ and $g_4$,
{\em i.e.}  $(2g_2 + 6g_4)\sum_i n_i b_i^\dg b_i $ from $\Ham_p$ and
$(g_2 + 6g_4)\sum_i n_i (b_i^{\dg,2} + b_i^2)$ from $\Ham_2$. Starting
from $g_2 = 0$ and making it increasingly more negative will at first
decrease these coefficients, thereby renormalizing the ground state
less. For even more negative $g_2$, however, $Z$ decreases
sharply as the absolute value of these coefficients increases; this is
paralleled by a strong increase in $N_{\text{ph}}$. Based on this
argument, the peak in $Z$ should occur for $-6g_4< g_2< -3g_4$, which
is indeed the case. The strong-coupling limit of a small polaron
(corresponding to small $Z$, large
$N_{\text{ph}}$ values) is therefore reached  either by  increasing
$|g_2|$ or by lowering $g_4$.

While this allows us to conclude that in the atomic limit the
crossover into the small polaron regime occurs at $\frac{g_2}{3g_4}
\approx -1.5$, it also illustrates the difficulty in defining an
effective coupling for this model. For the Holstein model, the
dimensionless effective coupling $\lambda$ is the ratio between the
ground-state energies in the atomic limit and in the free electron
limit; the crossover to the small polaron regime occurs at $\lambda
\sim 1$. For the double-well model the introduction of an effective
coupling is not as straightforward, because the atomic limit has
vastly different properties depending on the ratio $g_2/g_4$, so
comparing the energy in this limit to that of a free electron is not
sufficient. (Moreover, there is no analytic expression for the ground
state energy of the double well potential). For these
reasons, we continue to use the \emph{bare} coupling parameters $g_2$
and $g_4$ to characterize our model.

\begin{figure}[t]
\begin{center}
\includegraphics[width=0.5\textwidth]{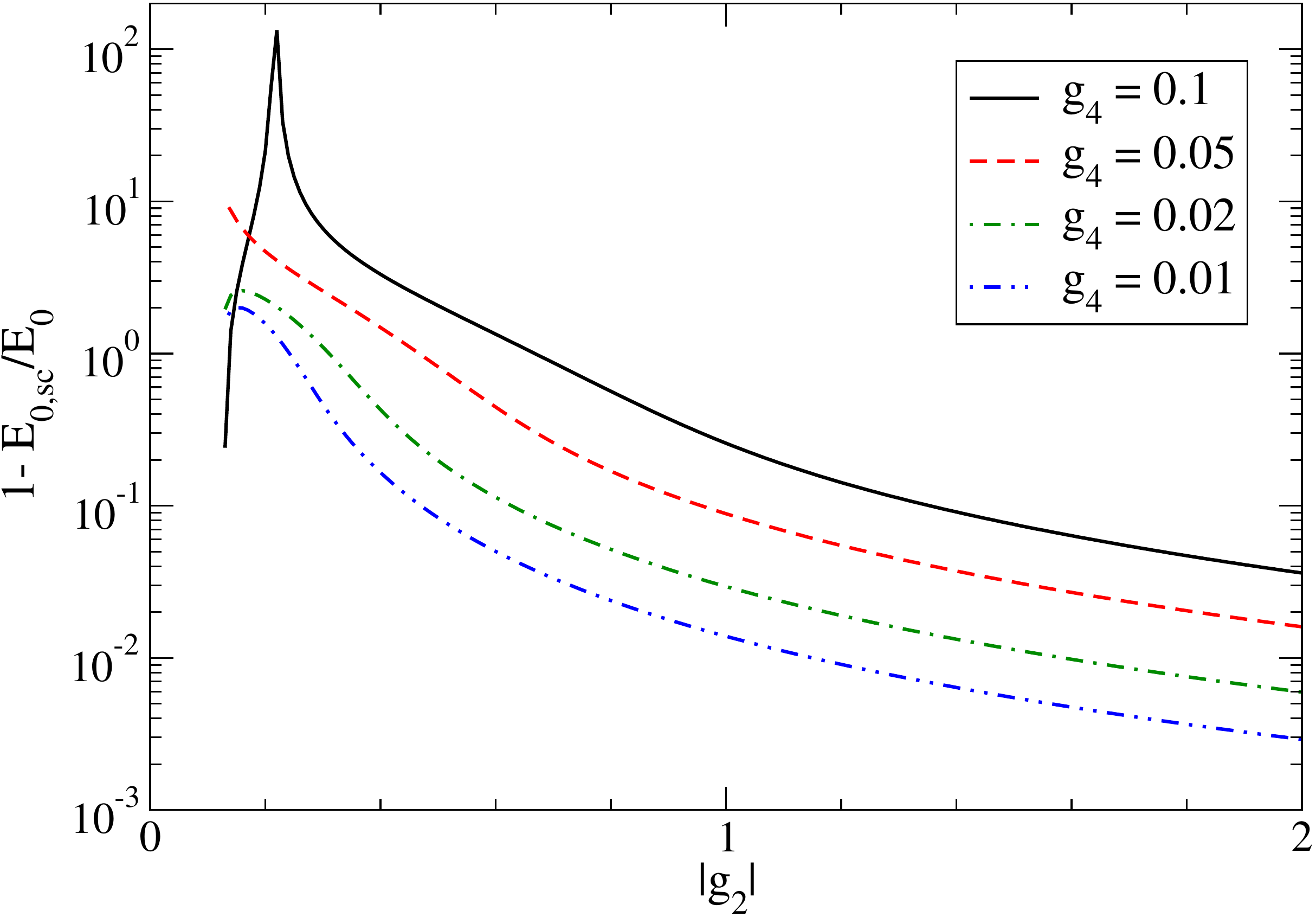}
\caption{\label{fig:atomic_limit_E}(color online) Relative error in
  the ground state energy when computed in the semiclassical
  approximation (see text for details). The coupling $g_2< -\Omega/4$
  is restricted to values
  for which there  is a double-well potential. Other parameters are
  like in Fig.~\ref{fig:atomic_limit_polaron}.}
\end{center}
\end{figure}

For strong coupling, we can accurately estimate the ground state energy
by using the barrier depth and effective harmonic frequency of the
double-well potential,
$E_{0,\text{sc}}= V_c(x_{\text{eq}}) +  \Omega_{\text{eff}}/2$.
Fig.~\ref{fig:atomic_limit_E} shows the relative
error of this estimate, which indeed decreases as parameters move
deeper into the small polaron regime. Since here the tunnelling between
the two wells also becomes increasingly smaller, one may think that we can
describe this regime accurately by assuming that the carrier becomes
localized in one of the wells (thus breaking parity), {\em i.e.} that we
can approximate the full lattice potential as being a single harmonic well
centered at either $x_{\text{eq}}$ or $-x_{\text{eq}}$. Of course,
the latter situation can be modelled with a linear Holstein model.

It turns out that this is not the case. In the standard Holstein model, the charge carrier cannot change the curvature of the lattice potential and thus cannot account for the difference between $\Omega$ and $\Omega_{\text{eff}}$.
To account for the change in the curvature of the well, one would have to consider at least a Holstein model with both linear and quadratic e-ph coupling terms.
Although it is possible to find effective parameters
$g_{1,\text{eff}}$, $g_{2,\text{eff}}$ and $\Omega_{\text{eff}}$
so that the resulting
lattice potential in the presence of the carrier has the same location
and curvature as one of the wells of the double-well potential, the
corresponding quasi-particle weight $Z_{\text{eff}}$ severely
underestimates $Z$. This is because the single well approximation
severely overestimates the lattice potential at $x = 0$, thereby
reducing the overlap between the ground state of the shifted well and
that of the original well. We conclude that the double-well coupling
cannot be accurately described by a (renormalized)  Holstein coupling
even in this simplest limit.

\subsection{Finite Hopping}

We focus  on results from the \emph{even} sector because it
describes  states accessible by injecting the carrier in the undoped
ground-state.  The odd sector is accessed only if the carrier is
injected into an excited state with
an odd number of phonons  present in the undoped system; we
briefly discuss this case at the end of the section.

\begin{figure}[t]
\includegraphics[width=0.48\textwidth]{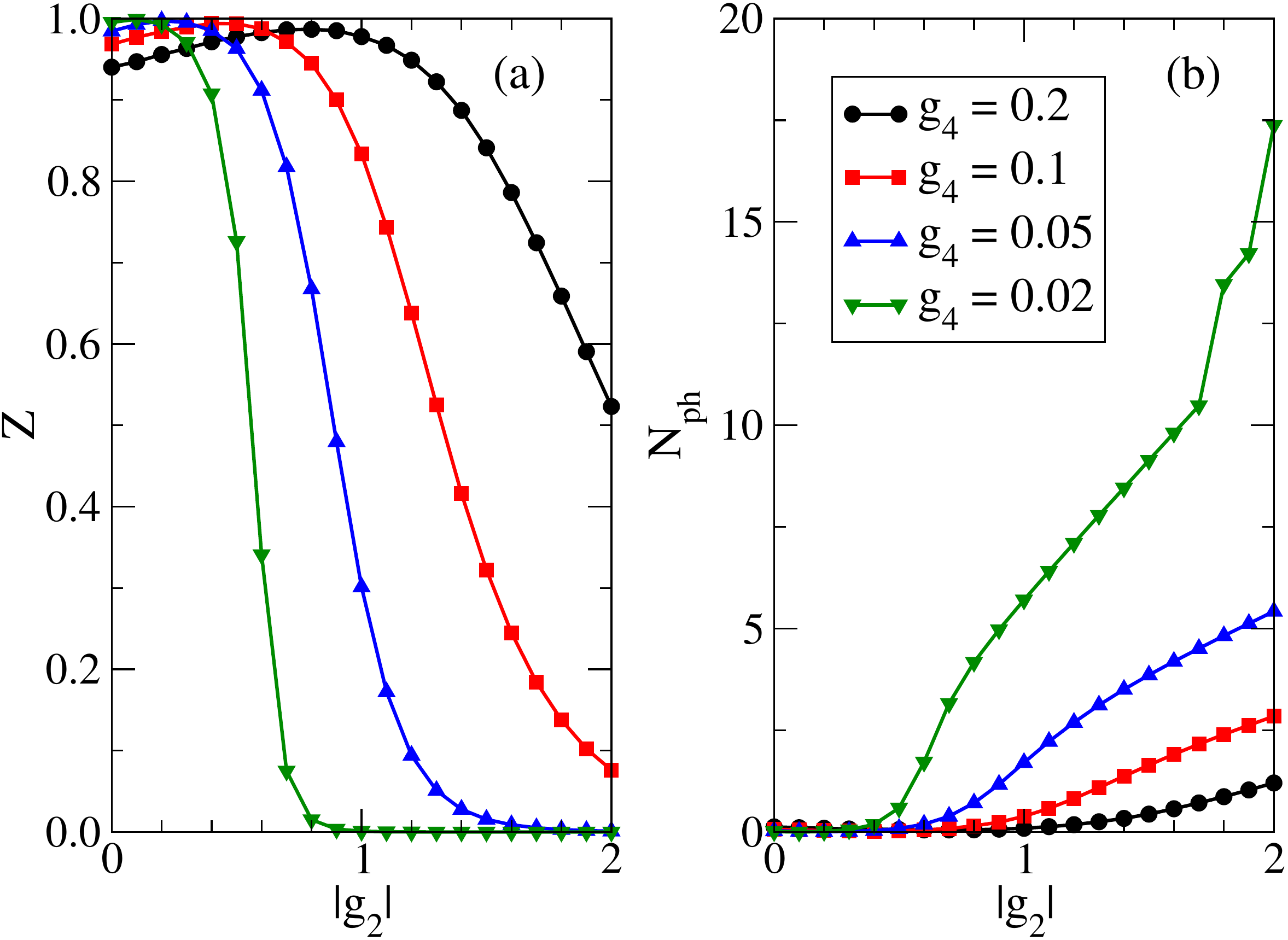}
  \caption{(color online) Polaron ground-state properties  in one dimension
 for various values of the
    quartic coupling term $g_4$ as a function of the quadratic coupling
    $g_2$: a)  quasiparticle weight, and b)  average number of
    phonons in the phonon cloud. Other parameters are $t = 1$, $\Omega
    = 0.5t$.  \label{fig:finite_t_polaron}}
\end{figure}

\begin{figure}[t]
\includegraphics[width=0.48\textwidth]{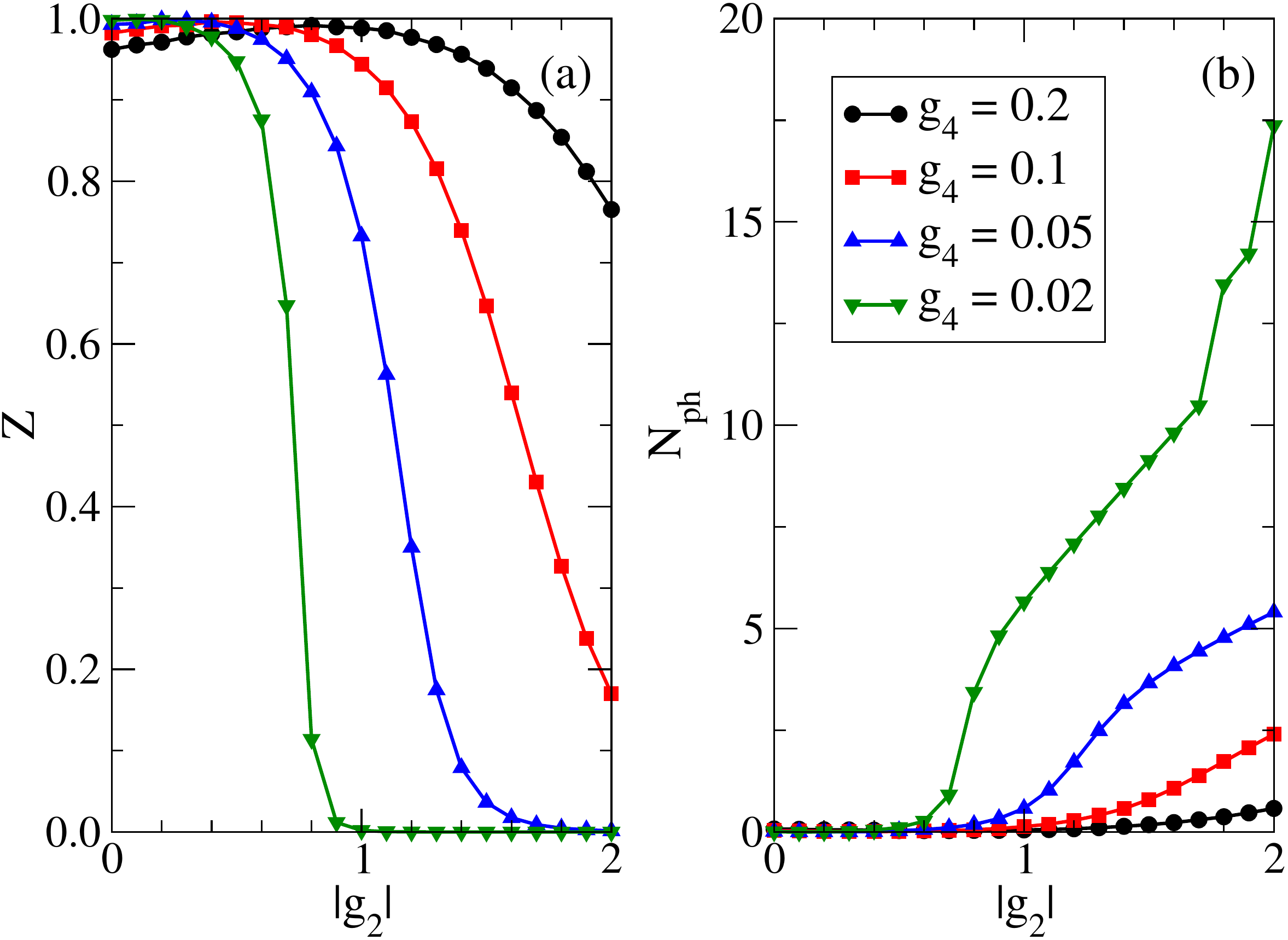}
  \caption{(color online) Polaron ground-state properties in two
    dimensions for various values of the quartic coupling term $g_4$
    as a function of the quadratic coupling $g_2$: a) quasiparticle
    weight, and b) average number of phonons in the phonon
    cloud. Other parameters are $t = 1$, $\Omega = 0.5t$.  }
  \label{fig:finite_t_polaron_2d}
\end{figure}

We begin by plotting the ground-state values of $Z$ and
$N_{\text{ph}}$, for 1D and 2D lattices, in
Figs. ~\ref{fig:finite_t_polaron} and \ref{fig:finite_t_polaron_2d}
respectively.  Since the MA self-energy is local, the effective
polaron mass
$m^*= m/Z$, where $m$ is the free carrier mass; we therefore do not
plot $m^*$ separately. Apart from $t=1$, the parameters are like in
Fig.~\ref{fig:atomic_limit_polaron}. Note that the kinks in the
$N_{\text{ph}}$ curves  for $g_4 = 0.02$ are not physical;
they arise from numerical difficulties in resolving the precise location of the
ground state peak  when $Z\rightarrow 0$.

Qualitatively, the polaron properties show the same dependence on
$g_2$ as in the atomic limit, but the shape and location of the
turnarounds is slightly modified: As one would expect, the presence of
finite hopping counteracts the formation of a robust polaron cloud and
increases the quasi-particle weight $Z$ for any given $g_2$ and $g_4$
when compared to the atomic limit.

The results in one and two dimensions are strikingly similar. The 2D
$Z$ is slightly larger than the 1D $Z$, and $N_{\text{ph}}$ in 2D is
slightly lower than in 1D.  This is expected because in higher
dimensions, the polaron formation energy is competing against a larger
 carrier kinetic energy. These results suggest that dimensionality is
 not playing a key role for the double-well model, similar to the
 situation for the Holstein model. This is why we did not consider 3D
 systems explicitly.

\begin{figure}[b]
\begin{center}
\subfigure[$g_2 = -0.5$]{\includegraphics[width=0.23\textwidth]{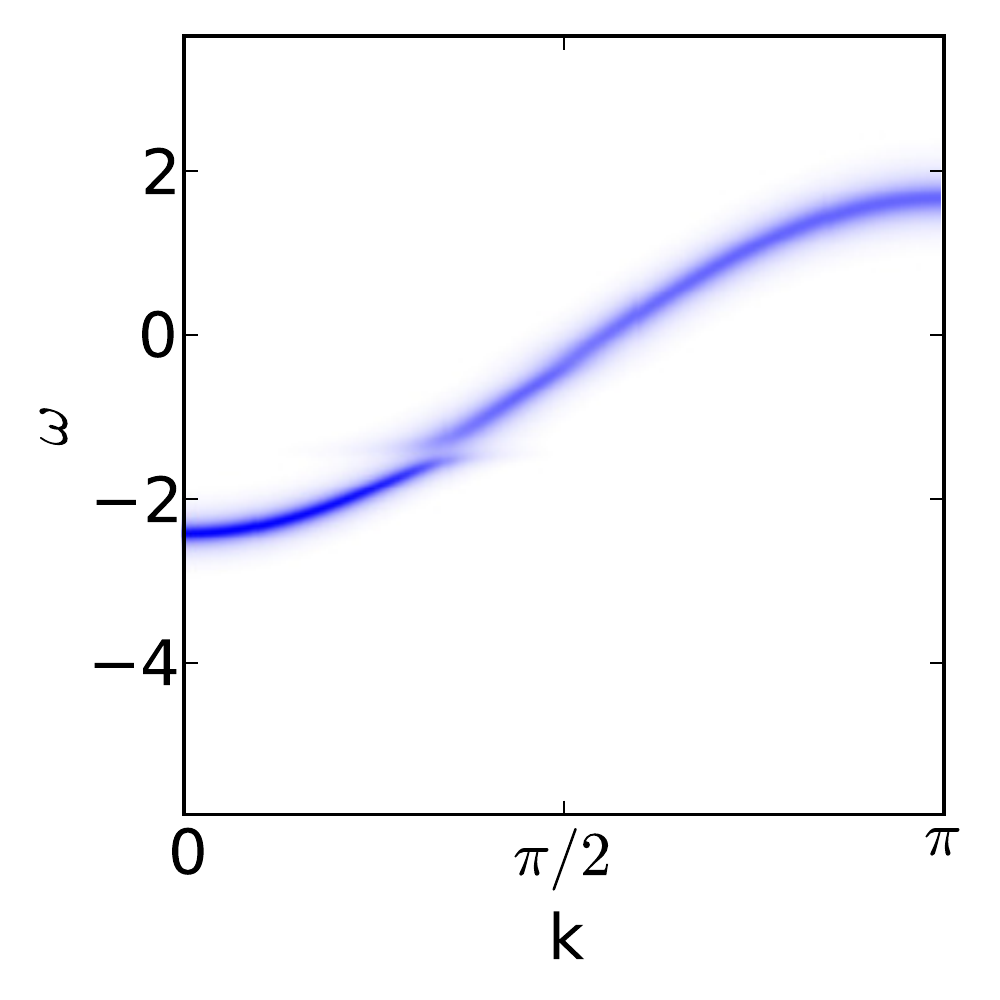}}
\subfigure[$g_2 = -1$]{\includegraphics[width=0.23\textwidth]{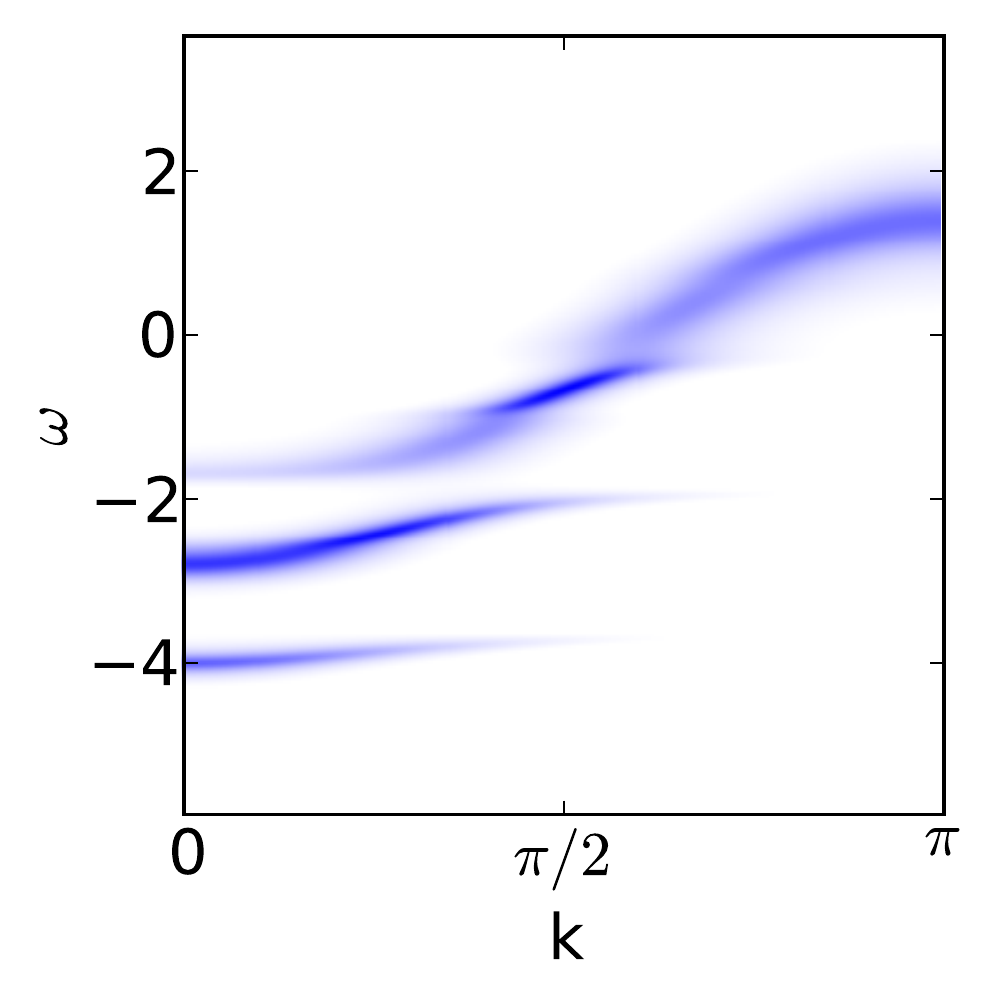}}
\subfigure[$g_2 = -1.5$]{\includegraphics[width=0.23\textwidth]{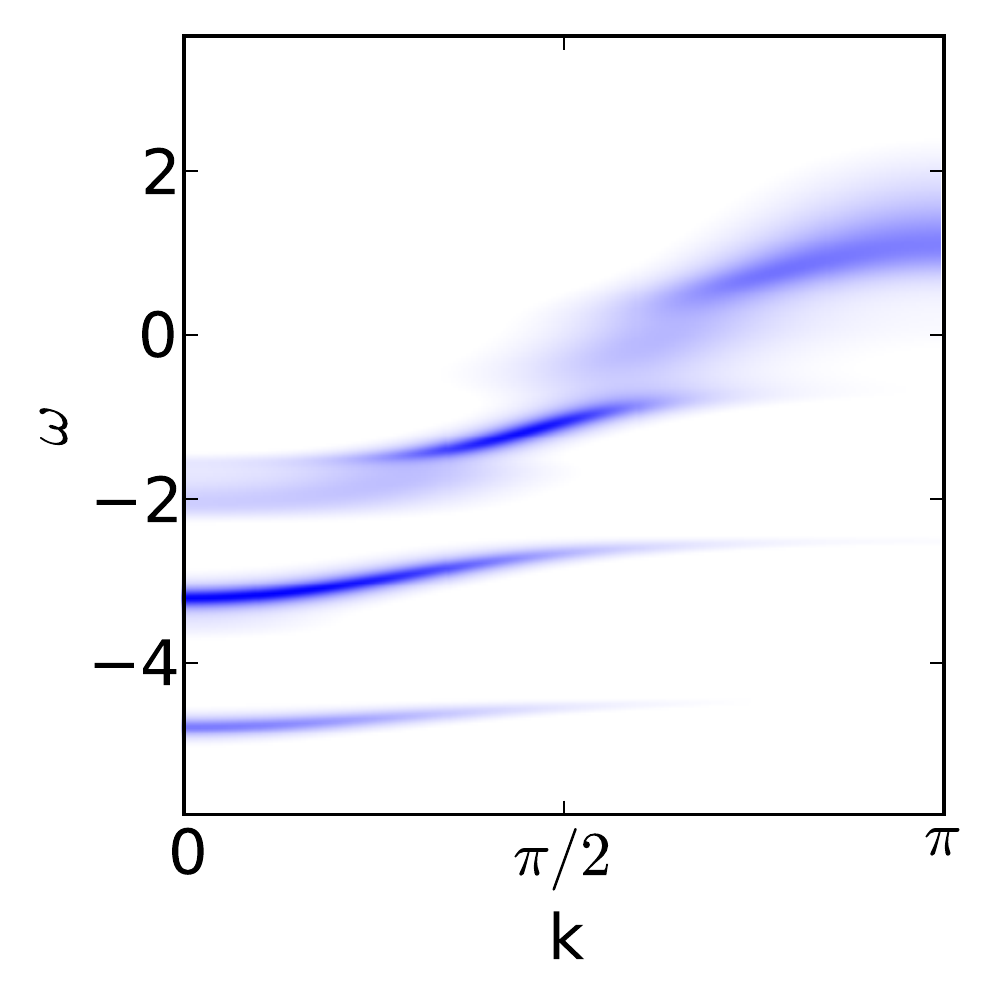}}
\subfigure[$g_2 = -2$]{\includegraphics[width=0.23\textwidth]{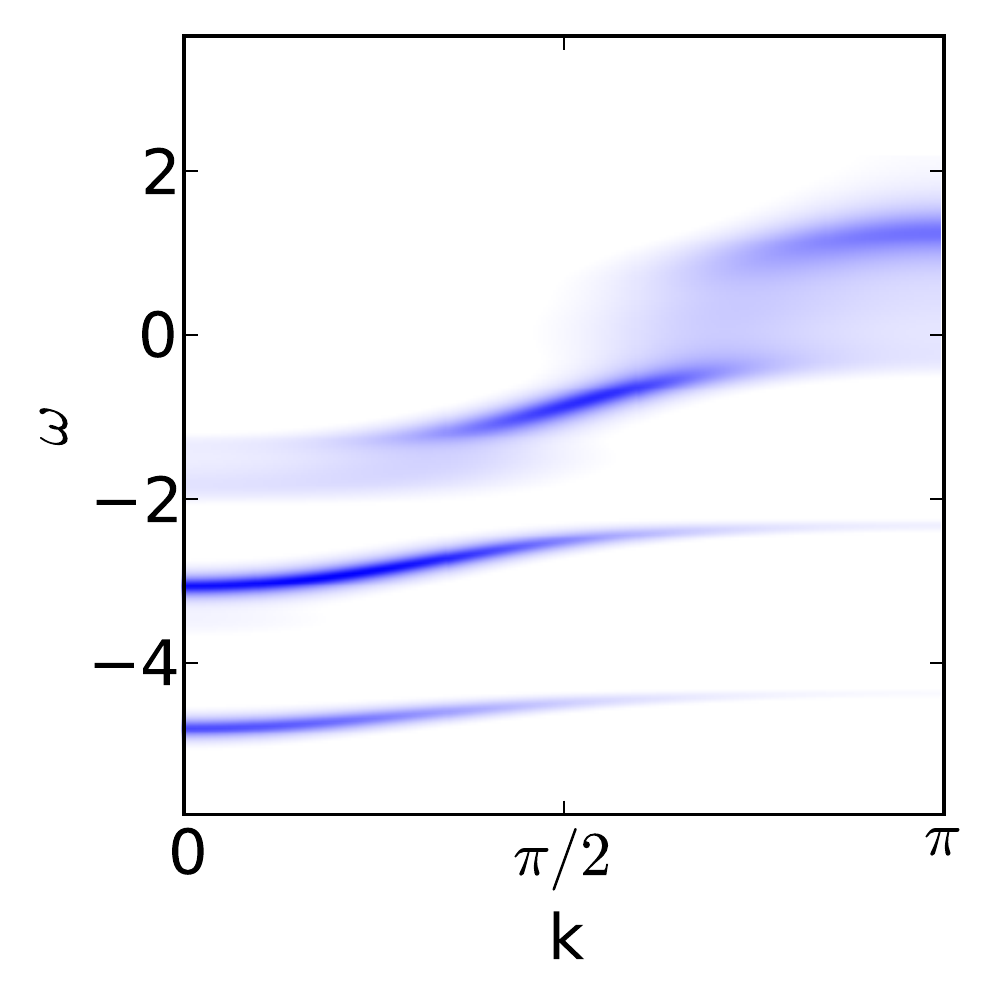}}
\caption{$A(k,\omega)$ in 1D, for $g_4 = 0.05$, $\Omega =
 0.5$ and $t = 1$, for various values of $g_2$.}
\label{fig:finite_t_spectrum}
\end{center}
\end{figure}

\begin{figure*}[t]
\begin{center}
\subfigure[ $g_2 = -0.5$]{\includegraphics[width=0.32\textwidth]{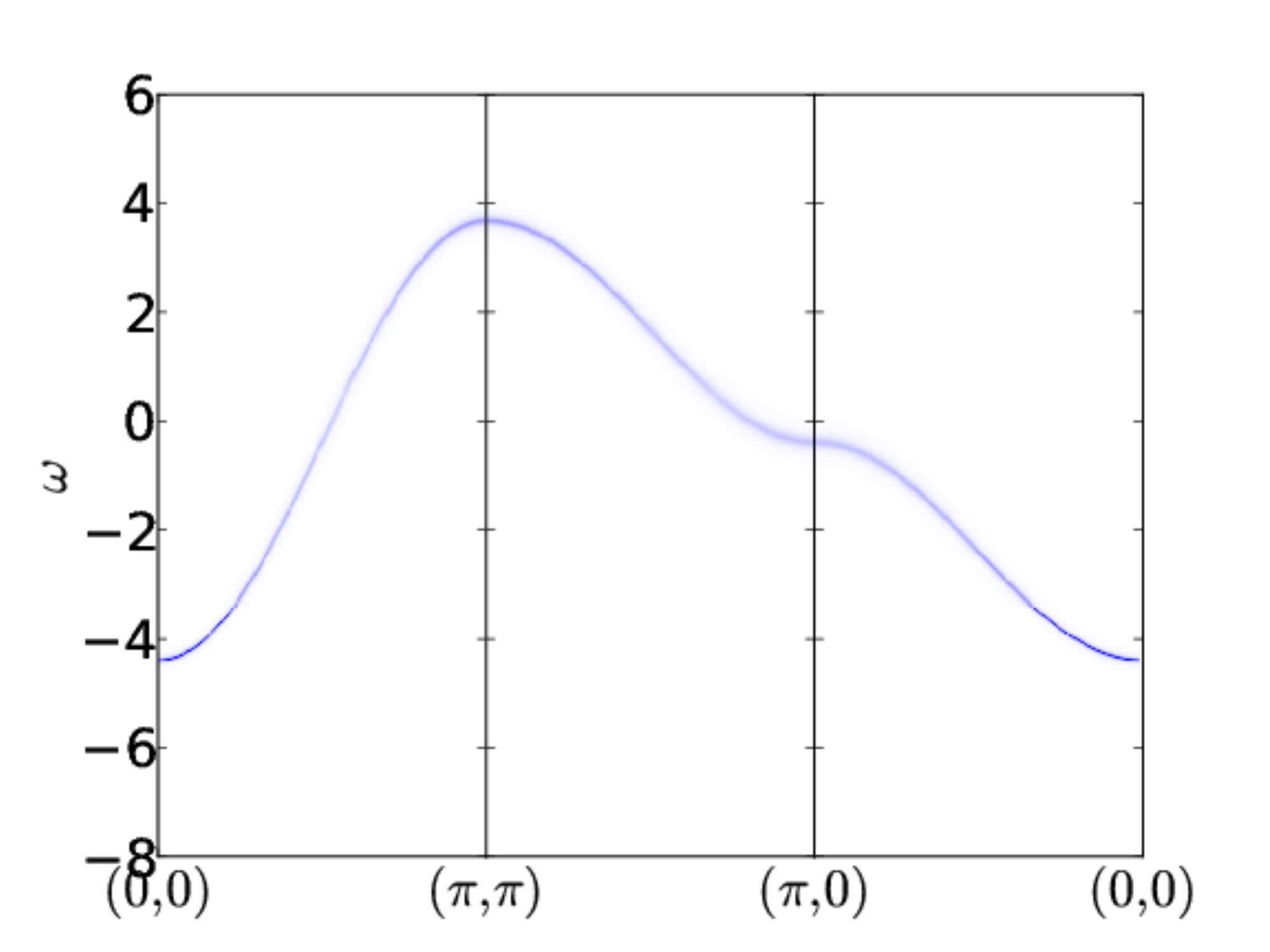}}
\subfigure[ $g_2 = -1$]{\includegraphics[width=0.32\textwidth]{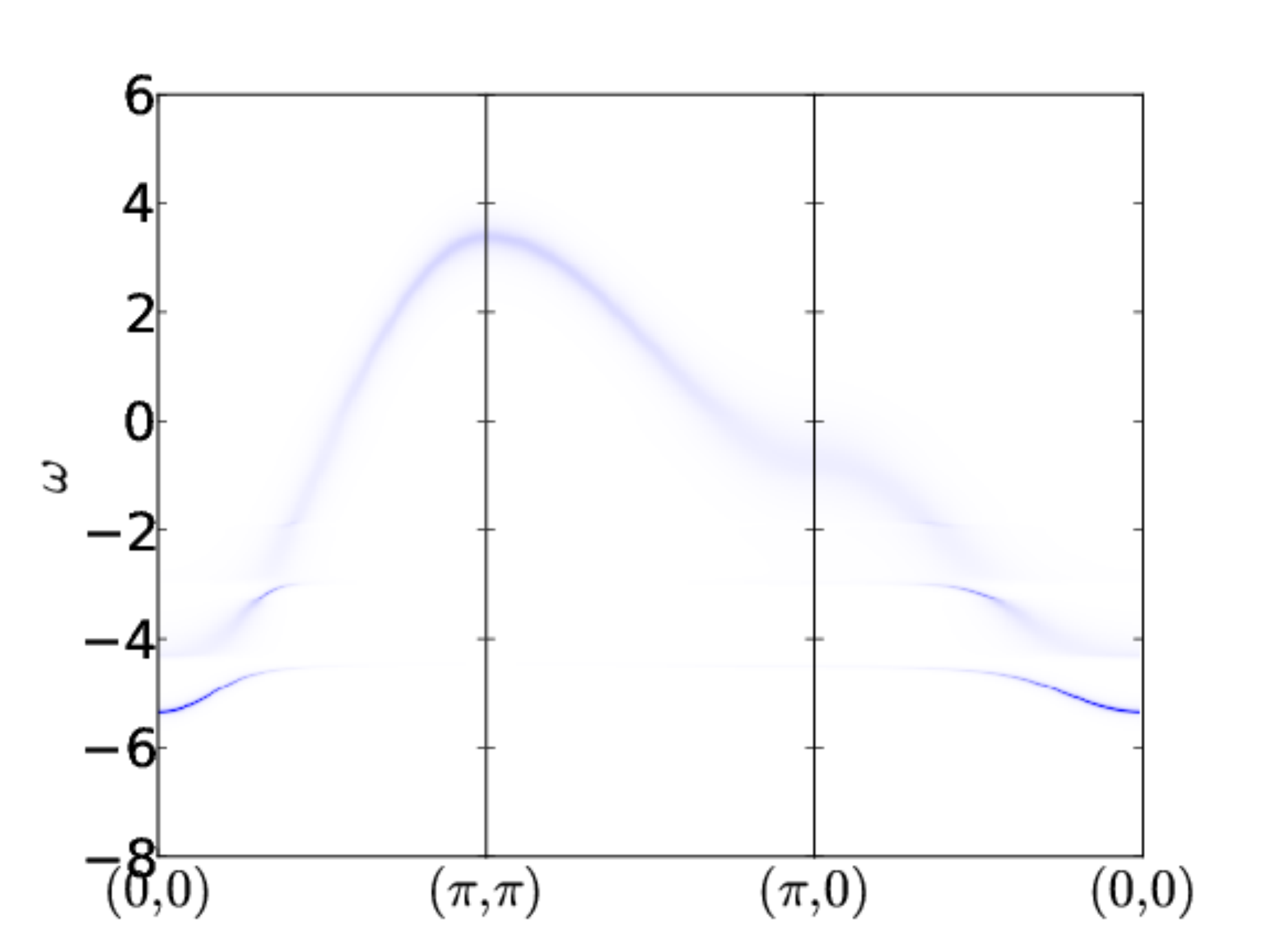}}
\subfigure[ $g_2 = -1.5$]{\includegraphics[width=0.32\textwidth]{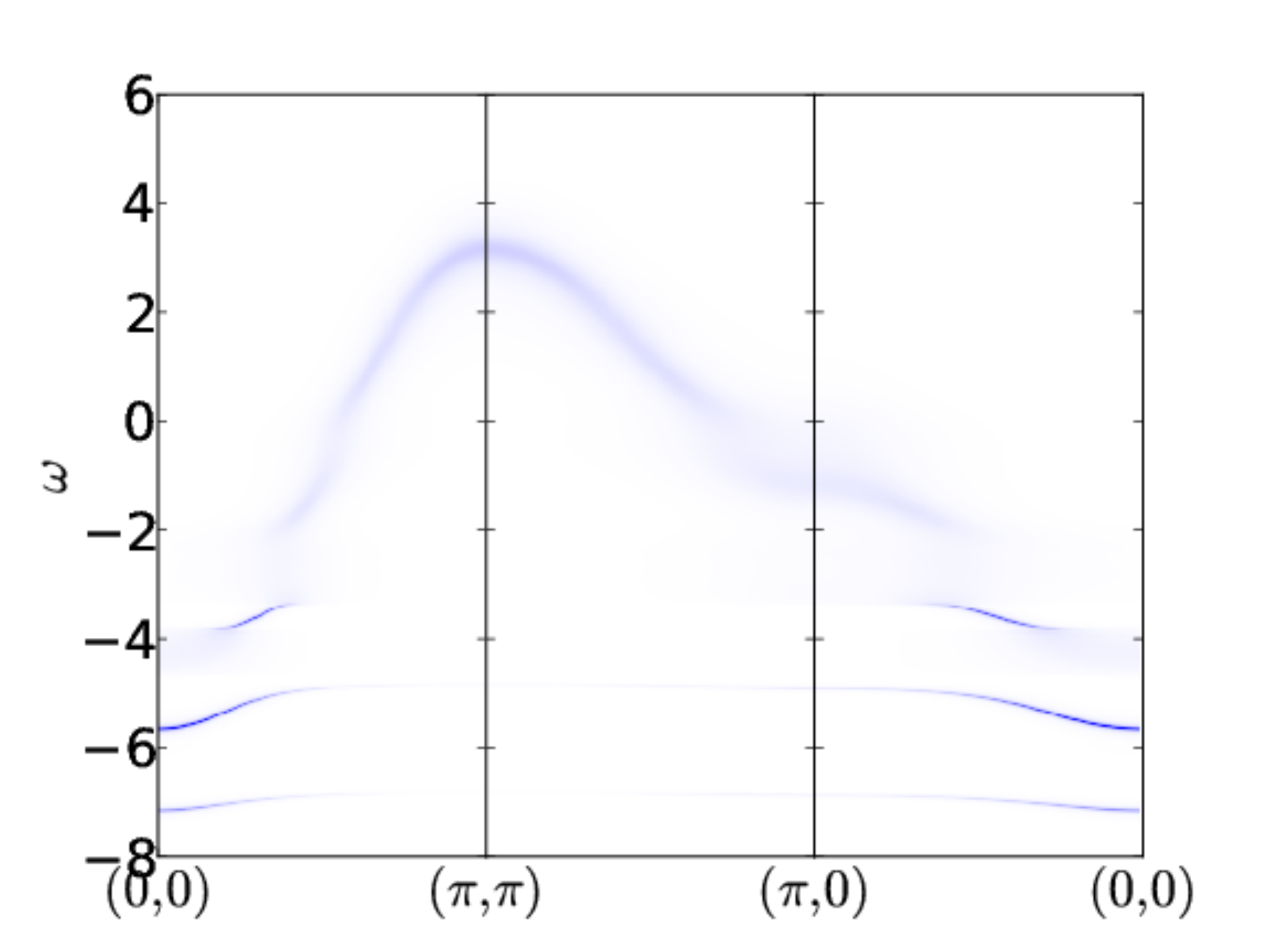}}
\caption{$A(k,\omega)$ in 2D, for $g_4 = 0.05$, $\Omega =
 0.5$ and $t = 1$, for various values of $g_2$.}
\label{fig:finite_t_spectral_2d}
\end{center}
\end{figure*}

We now move on to discuss the evolution of the spectral weight
$A(k,\omega) = -{1\over \pi} \text{Im} G(k,\omega)$ with increasing
$|g_2|$, at a fixed value of $g_4$.  This is shown in
Fig.~\ref{fig:finite_t_spectrum} for 1D, and in
Fig.~\ref{fig:finite_t_spectral_2d}  for 2D. Because the evolution is again
qualitatively similar in the two cases, we analyze in more
detail the 1D results.
Here, at small quadratic coupling $g_2 = -0.5$, we observe
the appearance of a polaron band  below a continuum of states. This continuum
begins at $E_0 + 2\Omega$, and consists of excited states comprising
the polaron plus two phonons far away from it.  (In our MA$^{(2)}$
approximation, the continuum actually begins at $E_0^{\text{MA}^{(0)}}
+ 2\Omega$, not at $E_0^{\text{MA}^{(2)}}
+ 2\Omega$, for reasons detailed in Ref.~\onlinecite{MA1}).

Note that due to the parity-preserving nature of the Hamiltonian
there is no analog of the polaron+one-phonon continuum starting at
$E_0 + \Omega$, which is observed in all linear coupling
models. Trying to mimic the results of the double-well coupling with
a linear model will, therefore, lead to a wrong assignment for the
value of $\Omega$.

At small $|g_2|$, the polaron band flattens out just below the
polaron+two-phonon continuum. With increasing $|g_2|$, its bandwidth
decreases as the polaron becomes heavier, and additional
bound states appear below the continuum. This is similar to the
evolution of the spectrum of a Holstein polaron when moving towards stronger
effective coupling.\cite{MA_goodvin}
However, as already discussed, this does not
mean that the two Hamiltonians can be mapped onto one another.

\begin{figure}[b]
\subfigure[]{\includegraphics[width=0.48\textwidth]{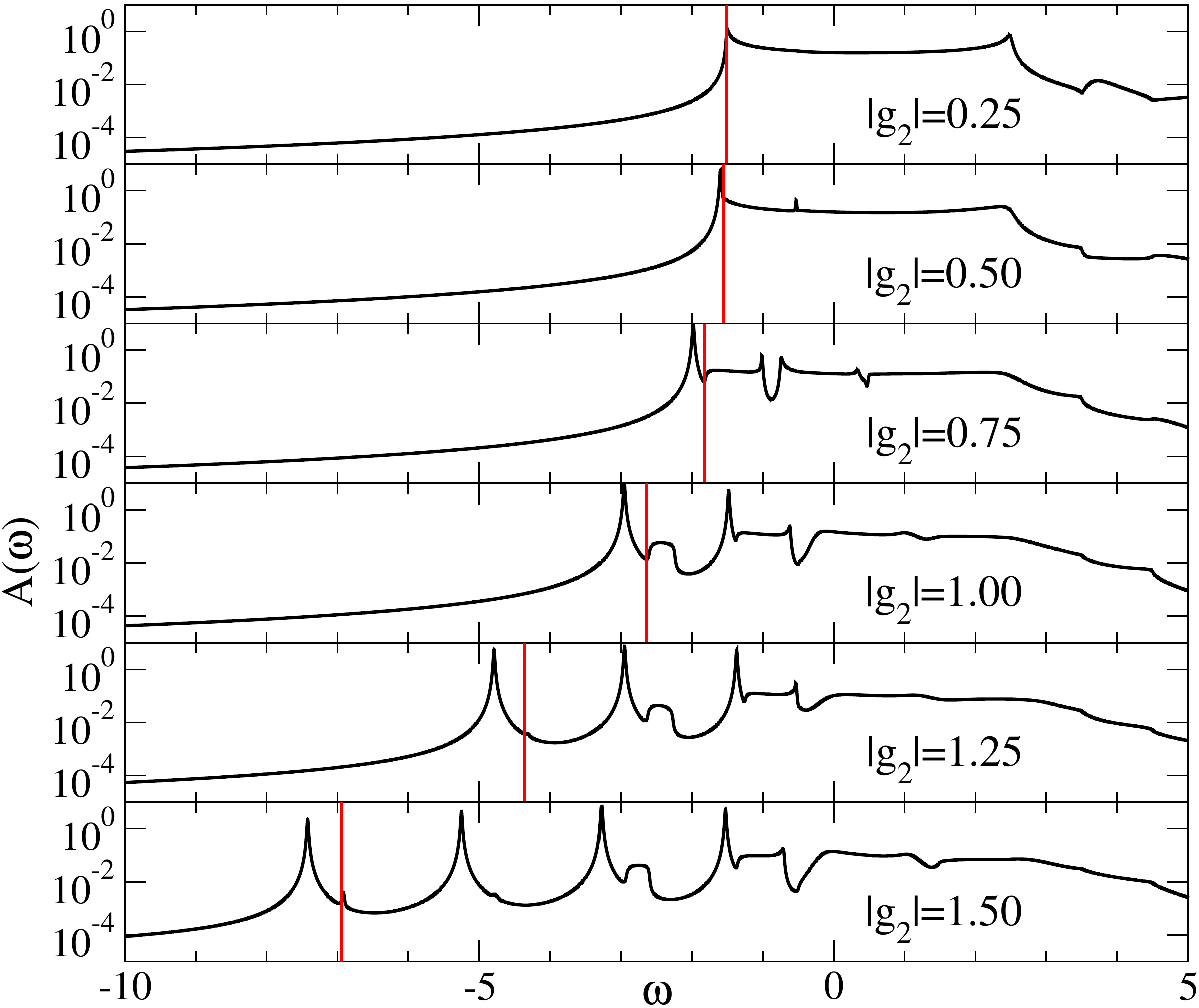}}
\caption{Real-space diagonal spectral function $A_{iii}(\omega)$ at $g_4 = 0.05$ for various values of (negative)
  $g_2$ in one dimension and for $\Omega = 0.5$. The $y$-axis has a
  logarithmic scale. The vertical bars indicate the position of
$E_0^{\text{even}} + \Omega$.}
\label{fig:finit_t_odd}
\end{figure}

For completeness, let us also discuss some of the features of the odd
sector.  In particular, we focus on the local Green's function
$G_{iii}(\omega)$, which can be written as
\begin{equation*}
G_{iii}(\omega) = \bar g_0(\tilde \omega) + \frac{\bar g_0(\tilde
  \omega)^2 (a^= - a^{\neq})}{1 - \bar g_0(\tilde \omega)(a^= -
  a^{\neq})}
\end{equation*}
with $\tilde \omega = \omega - \Omega - a^{\neq}$. One can verify that
$a^{\neq}$ equals the MA$^{(0)}$ self-energy for the \emph{even}
sector, up to a shift by $\Omega$ of its frequency.  The equation for
$G_{iii}(\omega)$ then shows that the odd sector spectral function
comprises two parts: (i) the first term is just
the momentum-averaged spectral function of the even-sector, shifted
in energy by $\Omega$ due to the presence of the extra phonon.  One
can think of these as states where the even-sector polaron does not
interact with the extra phonon. This contribution therefore has weight starting
from $E_0 + \Omega$; (ii) the second part describes
interactions between the polaron and the extra phonon. An interesting
question is whether these can lead to a bound state, {\em i.e.} to a
new polaron with odd numbers of phonons in its cloud.

This question is answered in Fig.~\ref{fig:finit_t_odd} where we plot
$A_{iii}(\omega)=-{1\over \pi} \text{Im} G_{iii}(\omega)$ for
different values of $|g_2|$ and $g_4=0.05$, $\Omega = 0.5$, $t = 1$,
in one dimension.  The vertical bars indicate the position of $E_0 +
\Omega$, where indeed a continuum begins, as expected from the
previous discussion. At sufficiently strong coupling $|g_2|$ we find a
discrete bound state below that continuum, showing that the polaron
can bind the extra phonon. In fact, it is more proper to say that the
extra phonon (which is localized somewhere on the lattice) binds the
polaron to itself and therefore localizes it. One can think of this as
an example of ``self-trapping'', except here there is an external
trapping agent in the form of the extra phonon.

One might wonder whether this localized bound state in the odd sector
could ever be at an energy below the polaron ground-state energy $E_0$
of the even
sector, {\em i.e.} become the true ground-state. This is not the case;
as explained above, in the atomic limit
the ionic states alternate between even and odd symmetry. Introducing
a finite hopping allows the polaron to further lower its energy by
delocalizing, but
this is only possible in the even sector. Thus, we always expect the
even-sector polaron to have an energy below that of this localized state.

As stated before, the
two subspaces with even and odd phonon number are never mixed, at least
at zero temperature. At finite temperature, the extra charge is
inserted not into the phonon vacuum but into a mixed state containing
a number of thermally excited phonons. We therefore expect the resulting
spectral function to show features of both the even and odd
sectors. To be more precise, some spectral weight should be shifted from the
even-sector spectral weight to the odd-sector spectral weight as $T$
increases and there is a higher probability to find one or more
thermal phonons in the undoped state. We plan to study the temperature
depend properties of this double-well coupling elsewhere.

\section{Summary and Discussions}\label{sec:summary}

Here we introduced and motivated a model for purely
quadratic e-ph coupling, relevant for certain types of intercalated
lattices, wherein the carrier dynamically changes the on-site lattice
potential from a single well  into a double well
potential. All the approximations made in deriving this model were
analyzed. In particular, we argued that ignoring the anharmonic
lattice terms at the sites not hosting the carrier should be a good
approximation. However, a more in-depth numerical analysis might be
needed to further validate this assumption.

We used the momentum average approximation to obtain the model's
ground state properties and its spectral function in the single
polaron limit, in one and two dimensions. We found that for
sufficiently strong quadratic coupling a small polaron forms. Although
the polaron behaves somewhat similarly to the polaron of the linear
Holstein model, the double-well model cannot be mapped onto an
effective linear model: apart from the difference in the location of
the continuum in the even sector, the double-well model also has an
odd sector that should be visible at finite $T$, and which is entirely
absent in the Holstein model. This is due to the double-well potential
model's invariance to local inversions of the ionic coordinate; this
symmetry is not found in the Holstein model. The polaron
in this odd sector is also qualitatively different from the Holstein
polaron, in that it is localized near the additional phonon present in
the system when the carrier is injected. Of course, if the assumption
of an Einstein mode is relaxed, then the phonon acquires a finite
speed and this polaron would become delocalized, as expected for a
system invariant to translations. However, this would still be
qualitatively different than a regular polaronic solution because this polaron's
dispersion would be primarily controlled by the phonon bandwidth, not
the carrier hopping.

Our results suggest that researchers interpreting their measurements
from, \textit{e.g.}, angular-resolved photoemission spectroscopy, must
carefully consider the nature of their system's e-ph coupling: if they
assume linear coupling where the lattice symmetry calls for a
quadratic one, the parameters extracted from fitting to such models
will have wrong values.

While we have laid here the basis for a thorough investigation of the
properties of the double-well e-ph coupling model, much work
remains. We believe that adjusting already existing numerical schemes
such as diagrammatic Monte Carlo to this model is straightforward and
look forward to a comparison of numerically exact results with our MA
results. In addition, there are certain ranges of parameters for which
MA is not well-suited, such as the adiabatic limit $\Omega \rightarrow
0$ at weak coupling, or systems with finite carrier densities. We
anticipate that these regimes will be explored with a range of
numerical and analytical tools, especially the finite carrier regime
which should be relevant for modelling ferroelectric materials.

We plan to extend our study of the double-well e-ph coupling beyond
the single-polaron limit. We deem especially interesting the parameter
range where the lattice potential remains a single well if only one
carrier is present, but changes into a double well when a second
charge is added. In this case, we anticipate the appearance of a
strongly bound bipolaron while the single polarons are relatively
light. Such states are not possible in the Holstein model.

Finally, extending our MA treatment to finite temperature should yield
interesting insights into the interplay between the two symmetry
sectors revealed by the spectral weight.

\acknowledgments

We thank NSERC and QMI for financial support.
\appendix
\section{Details for the even-sector}
\subsection{Coupling matrices}
\label{sec:appendix_even_matrices}

The matrices appearing in Eq. (\ref{w}) are:
\begin{align*}
  \gamma_n^=|_{11} &= 1 - \bar g_0(\omega - 4n\Omega) (8ng_2 + 24ng_4 + 96n^2
  g_4) \\
  \gamma_n^=|_{12} &= - \bar g_0(\omega - 4n\Omega)(g_2 + 6g_4 + 16ng_4) \\
  \gamma_n^=|_{21} &= - \bar g_0(\omega - (4n+2)\Omega)\big(
    (g_2+6g_4)(4n+2)^{\bar 2} + \\ & \phantom{=} 4g_4(4n+2)^{\bar 3}\big) \\
  \gamma_n^=|_{22} &= 1 - \bar g_0(\omega - (4n+2)\Omega)\big(
    (8n+4)g_2 + \\ &\phantom{=} (24n+12)g_4 + 24(2n+1)^2 g_4\big)
\end{align*}
\begin{align*}
  \alpha_n^=|_{11} &= \bar g_0(\omega - 4n\Omega) (g_4 (4n)^{\bar 4}) \\
  \alpha_n^=|_{12} &= \bar g_0(\omega - 4n\Omega) \left((g_2+6g_4)(4n)^{\bar 2} +
  4g_4(4n)^{\bar 3}\right) \\
  \alpha_n^=|_{21} &= 0 \\
  \alpha_n^=|_{22} &= \bar g_0(\omega - (4n+2)\Omega \left( g_4 (4n+2)^{\bar 4}\right)
\end{align*}
\begin{align*}
  \beta_n^=|_{11} &= \bar g_0(\omega - 4n\Omega) g_4 \\
  \beta_n^=|_{12} &= 0 \\
  \beta_n^=|_{21} &= \bar g_0(\omega - (4n+2)\Omega)(g_2 + 6g_4 + (16n+8) g_4) \\
  \beta_n^=|_{22} &=  \bar g_0(\omega - (4n+2)\Omega) g_4
\end{align*}
The matrices for $\neq$ sector are the same if we substitute $n
\rightarrow n - 1/2$ everywhere except in the argument of $\bar g_0(\omega)$.

\subsection{Manipulation of the EOMs}
\label{sec:appendix_even_EOMs}
We can rewrite the EOM of $F_1$ by inserting the matrices
$A_1^=$ and $A_1^{\neq}$ and
collecting terms. This results in
\begin{multline}
  F_1(ij) = G_0(j-i,\omega-2\Omega)\left[ a_0^= G(j) + a_1^= F_1^=(j)\right] \\
 + \sum_{l\neq j} G_0(l-i,\omega-2\Omega) a^{\neq} F_1^{\neq}(lj).
\end{multline}
where we omit the arguments $k$ and $\omega$ for shorter notation. We give
expressions for the various coefficients below. For now, we rewrite
the EOM as
\begin{multline}
  F_1(ij) = G_0(j-i,\omega-2\Omega) \\ \times \left[ a_0^= G(j) + (a_1^= -
    a_1^{\neq})F_1^=(j)\right] \\
  + \sum_l G_0(l-i,\omega-2\Omega) a^{\neq} F_1(lj).
\end{multline}
Defining $G_0(\omega)_{ij} := G_0(j-i, \omega)$, we can write this as a matrix
product:
\begin{multline*}
  \sum_l \left[ \delta_{il} - a^{\neq} G_0(\omega-2\Omega)_{il}\right]
  F_1(lj) = \\
  G_0(\omega-2\Omega)_{ij} \left[ a_0^= G(j) + (a_1^= - a_1^{\neq}) F_1(jj)\right].
\end{multline*}
We multiply this  from the left with $G_0^{-1}(\omega - 2\Omega)$
and obtain
\begin{multline*}
  \sum_l \left[ G_0^{-1}(\omega - 2\Omega)_{rl} - a^{\neq} \delta_{rl} \right]
  F_1(lj) = \\ \delta_{rj} \left[ a_0^{\neq} G(j) + (a_1^= - a^{\neq}) F_1(jj)\right].
\end{multline*}
Next, we use the fact that $G_0^{-1}(\omega - 2\Omega)_{rl} =
\delta_{rl}(\omega - 2\Omega)  -
\hat H_{rl}$, so subtracting $a^{\neq}\delta_{rl}$ from this just shifts its
frequency to obtain
$G_0^{-1}(\omega - 2\Omega - a^{\neq})_{rl}$. As a result:
\begin{multline*}
  F_1(ij) = G_0(\omega - 2\Omega - a^{\neq})_{ij}
  \\ \times \left[ a_0^= G(j) +
    (a_1^= - a_1^{\neq}) F_1(jj)\right].
\end{multline*}
Since in the EOM for $G$ we only require $F_1(jj)$, we solve for that
diagonal element and obtain
\begin{equation*}
  F_1(jj) = \frac{\bar g_0(\omega - 2\Omega - a^{\neq}) a_0^= G(j)}{1 - \bar
    g_0(\omega - 2\Omega - a^{\neq})(a_1^= - a_1^{\neq})}.
\end{equation*}
The coefficients are obtained by just inserting the appropriate matrices $A_n$ into
the EOM and collecting terms:
 \begin{align*}
  a_0^= &= 2g_2 + 12g_4 + (g_2 + 14g_4)A_1^=|_{11} + g_4A_1^=|_{21}  \\
  a_1^= &= 4g_2 + 36g_4 + (g_2 + 14g_4)A_2^=|_{12} + g_4A_2^=|_{22} \\
  a^{\neq} &= (g_2 + 6g_4) A_1^{\neq}|_{12} + g_4 A_1^{\neq}|_{22}
\end{align*}
Finally, $F_1(jj)$ are used in Eq. (\ref{geq})  to obtain
$G(k,\omega)$.

\section{Details for the odd-sector}
\label{sec:appendix_odd}
\subsection{Equations of Motion}
Starting from the EOM for $G_{ijl}(\omega)$,
we let $\Ham_1$ act on the states in those sums, to find for the
diagonal state:
\begin{multline*}
  \Ham_1 c_l^\dg b_l^\dg \ket{0} =
  (2g_2 + 12g_4) c_l^\dg b_l^\dg \ket{0}
   \\ + (g_2 + 10 g_4) c_l^\dg b_l^{\dg, 3} \ket{0}  + g_4 c_l^\dg b_l^{\dg, 5} \ket{0}
\end{multline*}
while for the off-diagonal ones:
\begin{multline*}
  \Ham_1 c_{i'}^\dg b_l^\dg \ket{0} =
  (2g_2 + 6g_4) c_{i'}^\dg b_l^\dg \ket{0}
  \\ +(g_2 + 6g_4) c_{i'}^\dg b_l^{\dg, 2} b_l^\dg \ket{0} +
  g_4 c_{i'}^\dg b_l^{\dg, 4} b_l^\dg \ket{0}.
\end{multline*}
We now define the generalized Green functions as:
\begin{equation*}
  F_n(k,i,j,\omega) = \braket{k | \hat G(\omega) c_i b_i^{\dg, 2n} b_j | 0}
\end{equation*}
so we always have the extra phonon at site $j$.
The equation of motion for $G$ then becomes:
$ G_{ijl}(\omega) = G_{0}(j-i,\omega-\Omega) + \big[(2g_2 + 12g_4) F_0^=(l)  + (g_2 + 10 g_4)F_1^=(l) + g_4 F_2^=(l)\big] G_{ill}+
    \sum_{i'\not= l} \Big[(2g_2 + 6g_4)
    F_0^{\neq}(i',l) + (g_2 + 6g_4)F_1^{\neq}(i',l) + g_4 F_2^{\neq}(i',l) \Big] G_0(i'-i,\omega-\Omega).$
Again, we start by separating the cases $F_n^=$ and $F_n^{\neq}$. The resulting
equations of motion for $F_n^=$ are like those of the even-sector $F_n^=$ with $n
\rightarrow n + 1/2$, while those for $F_n^{\neq}$ are like those of the even-sector
$F_n^{\neq}$ with $n \rightarrow n+1$.

In the spirit of MA$^{(2)}$, only the EOM for $G$, which already has one phonon
present, is kept exact, while in the EOMs for all the $F_n$ with $n
\ge 1$ we approximate
$G_0(i-j,\omega)\rightarrow \delta_{ij}\bar g_0(\omega)$. We introduce
matrices $W_n = (F_{2n-1},
F_{2n})$. Again we obtain an equation like Eq. (\ref{w}), where now:
  \begin{align*}
    \gamma^=_{11} &= 1 - \bar g_0(\omega-(4n-1)\Omega)((4n-1)(2g_2+6g_4 \\ &\phantom{=} +
    6g_4(4n-1)) \\
    \gamma^=_{12} &= -\bar g_0(\omega-(4n-1)\Omega)\left(g_2+6g_4 + 4g_4(4n-1)\right)\\
    \gamma^=_{21} &= -\bar g_0(\omega-(4n+1)\Omega)\Big((4n+1)^{\bar
        2}(g_2+6g_4) \\ &\phantom{=} + (4n+1)^{\bar 3} \cdot 4g_4\Big) \\
    \gamma^=_{22} &= 1 - \bar g_0(\omega - (4n+1)\Omega) (4n+1) \\ &\phantom{=}\times\left(2g_2 +
      6g_4 + 6g_4(4n+1)\right)
  \end{align*}
  \begin{align*}
    \alpha^=_{11} &= \bar g_0(\omega - (4n-1)\Omega) (4n-1)^{\bar 4} g_4 \\
    \alpha^=_{12} &= \bar g_0(\omega - (4n-1)\Omega)
    	\\ &\phantom{=} \times \left(
      (4n-1)^{\bar 2} (g_2 + 6g_4) + (4n-1)^{\bar 3} \cdot 4g_4\right) \\
    \alpha^=_{21} &= 0 \\
    \alpha^=_{22} &= \bar g_0(\omega - (4n+1)\Omega) (4n+1)^{\bar 4} g_4
  \end{align*}
  \begin{align*}
    \beta^=_{11} &= \bar g_0(\omega - (4n-1)\Omega) g_4 \\
    \beta^=_{12} &= 0 \\
    \beta^=_{21} &= \bar g_0(\omega - (4n+1)\Omega) \left(g_2 + 6g_4 +
      4g_4(4n+1)\right) \\
    \beta^=_{22} &= \bar g_0(\omega - (4n+1)\Omega) g_4
  \end{align*}
The matrices for $W_n^{\neq}$ are obtained from these by replacing $n
\rightarrow n - 1/4$ everywhere
except in the argument of  $\bar g_0$. The remaining
steps are in close analogy to those for obtaining the even-sector
Green's function and not reproduced here.

The coefficients occurring in the final results for the odd-sector
Green's function are
\begin{align*}
a^=_{\text{o}} &= 2 g_2 + 12 g_4 + (g_2 + 10 g_4) A_1^=|_{1,2} + g_4  A_1^=|_{2,2} \\
a^{\neq}_{\text{o}} &= (g_2 + 6  g_4)  A_1^{\neq}|_{1,2} + g_4  A_1^{\neq}|_{2,2}.
\end{align*}

\subsection{Momentum space Green's functions}
Rather than having the phonon present at a lattice site $l$, we can
construct an electron-phonon state of total momentum $K$ as $\ket{K,n}
= \sum_i e^{iKR_i}/\sqrt{N} c_i^\dg b_{i+n}^\dg \ket{0}$ where $n$ is
the relative electron-phonon distance. It is easy to show that
$\braket{K, m | \hat G(\omega) | K, n} = G_{i,i+n-m,i+n}(\omega)
\exp(iKa(n-m))$ where $a$ is the lattice constant. In particular, the
odd-polaron propagator $n = m = 0$ is just the completely local real
space propagator $G_{iii}(\omega)$. In other words, the odd-sector
polaron shows no dispersion at all.

  Another Green's function of interest is given by
  \begin{equation*}
    \braket{k', q' | \hat G(\omega) | k, q} =
    \braket{0 | c_{k'} b_{q'} \hat G(\omega) b_q^\dg c_k^\dg | 0}
  \end{equation*}
  where we insert an electron of momentum $k$ into a system where the
  phonon has momentum $q$. Conservation of total momentum demands that
  $k + q = k' + q'$. It is again easy to show that the resulting
  propagator is
  \begin{multline*}
    \braket{k', q' | \hat G(\omega) | k, q} =
    \delta_{kk'}\delta{qq'} G_0(k, \tilde \omega) + \\ \frac{1}{N} G_0(k',
    \tilde \omega) G_0(k,\omega) \cdot \frac{a_o^= - a_0^{\neq}}{1 -
      \bar g_0(\tilde \omega) (a_o^= - a_0^{\neq})}.
  \end{multline*}
  Since the latter term vanishes in the thermodynamic limit $N
  \rightarrow \infty$, we are left with just the even-sector polaron
  propagator. This is to be expected: In an infinite system, an
  electron does not scatter off a single impurity. If instead we
  assume a finite but low density $n_p$ of phonons, the prefactor
  $1/N$ in the scattering term is replaced with $n_p$.

  This brief analysis shows that the interesting physics of the odd
  phonon number sector are best observed in real space.

\begin{figure}[t]
  \centering
\includegraphics[width=0.4\textwidth]{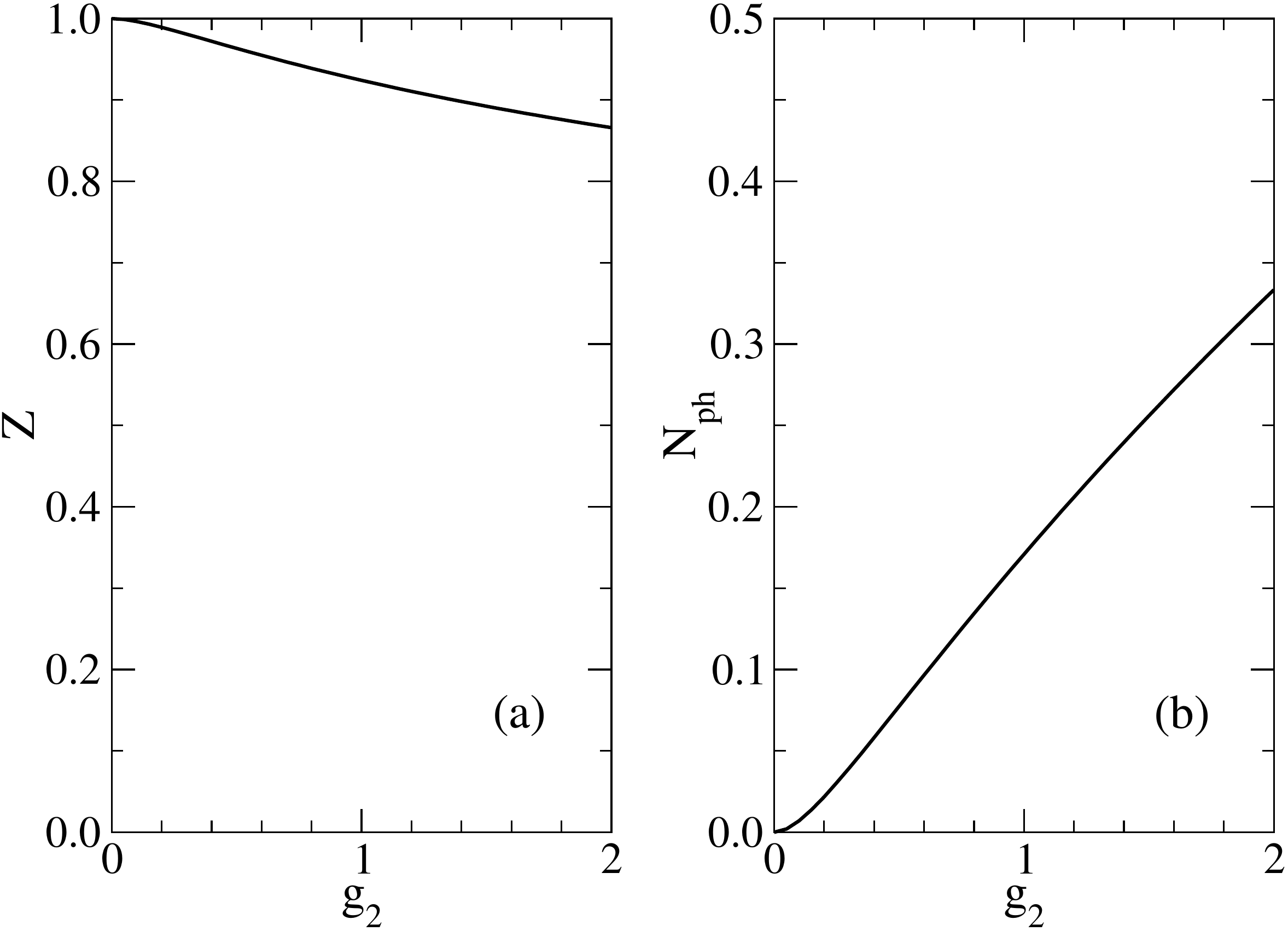}
  \caption{a) Quasiparticle weight $Z$, and (b)
    average number of phonons for a quadratic model with $g_2>0,
    g_4=0$ in the atomic limit $t=0$, for $\Omega=1$.}
\label{fig:quadratic_effects}
\end{figure}

\section{Quadratic e-ph coupling with $g_2>0$}
\label{newap}
\begin{figure}[tbhp]
  \centering
\includegraphics[width=0.45\textwidth]{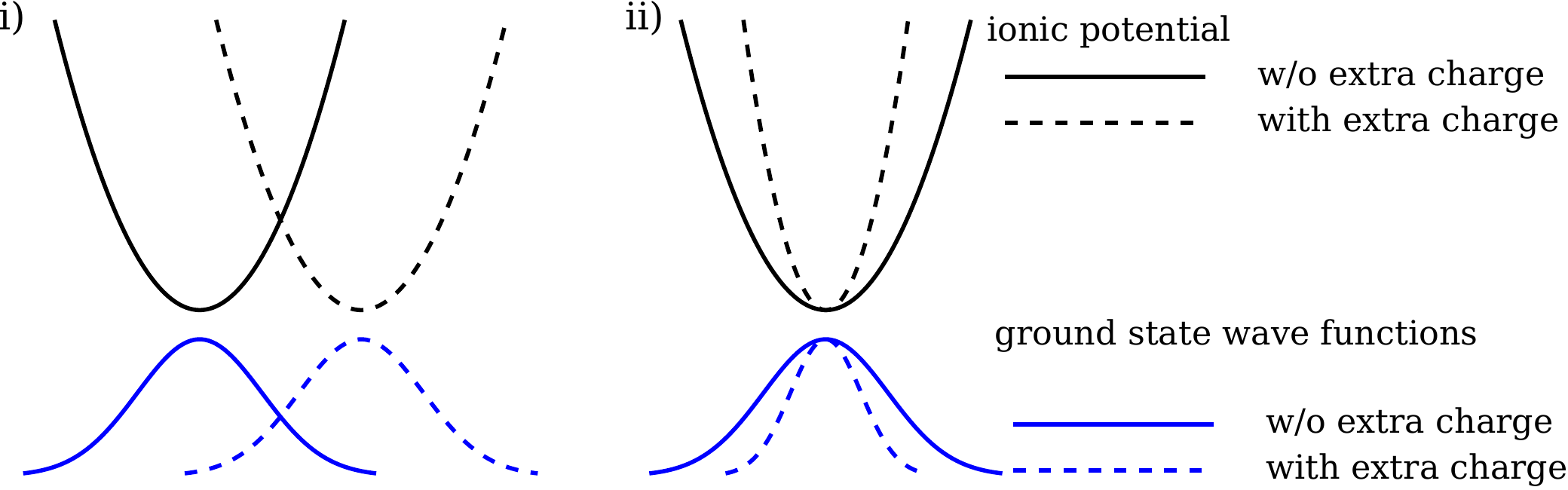}
\caption{(color online)
  Sketch of the lattice potential for i) Holstein, and ii) $g_2>0$
  quadratic  models. Full (dashed) lines indicate ionic potential and
  ground state wavefunction without (with) an extra charge on the
  site.}
\label{fig:lin_quad_overlap}
\end{figure}
Fig.~\ref{fig:quadratic_effects} shows that for $g_2 > 0$, $g_4 = 0$,
the e-ph coupling has an extremely weak effect even in the atomic
limit $t = 0$, since  the quasiparticle weight $Z$
remains very close to $1$ while the average number of phonons is
very small. An explanation for this behaviour is sketched in
Fig.~\ref{fig:lin_quad_overlap}: in the linear Holstein model, the
carrier displaces the harmonic lattice potential of its site, as
sketched in the left panel. The overlap between the ground state
wavefunctions of the original and the displaced potentials is then the
overlap between the tails of two Gaussians with different centers,
which decreases exponentially with increasing displacement. Indeed, in the
atomic limit for the linear Holstein model $Z \sim
\exp[-(g/\Omega)^2]$.  In the purely quadratic model with positive
$g_2$, however, the electron merely changes the shape of the well by
increasing $\Omega$ to $\Omegaat$. The overlap between the ground
states of the original and modified potential is that of two
Gaussians with the same center but different widths. We can calculate
this overlap analytically to find
\begin{equation}
  Z = \sqrt{1 - \left(\frac{\Omega - \Omegaat}{\Omega +
      \Omegaat^2}\right)^2}
\end{equation}
For $\Omega = 1.0$, even for $g_2 = 100\Omega$ we still have $Z
\approx 0.42$. We conclude that a positive, purely quadratic
electron-phonon coupling has negligible effect on the dynamics of a
charge carrier. In particular, no crossover into the small polaron
regime occurs for positive $g_2$ for any reasonable coupling
strength. Finite $t$ results (not shown) fully support this conclusion.

%

\end{document}